\documentclass[]{aa}
\usepackage{graphicx}
\usepackage{multirow}
\usepackage[varg]{txfonts}
\usepackage{xcolor}
\usepackage{booktabs}      
\newcommand{\kms}{km\,s$^{-1}$}

\usepackage{url}
\usepackage{relsize}

\usepackage{hyperref}
\begin{document}

\title{Elemental abundances\\ in the remnant of the ancient eruption of CK Vulpeculae}
\author{R. Tylenda\inst{\ref{inst1}},
        T. Kami\'nski\inst{\ref{inst2}}\thanks{Submillimeter Array Fellow}, 
        A. Mehner\inst{\ref{inst3}},
        } 
\institute{\centering 
    Nicolaus Copernicus Astronomical Center, Polish Academy of Sciences, Rabia{\'n}ska 8, 87-100 Toru\'n, 
         \email{tylenda@ncac.torun.pl} \label{inst1}
    \and Center for Astrophysics, Harvard \& Smithsonian, 60 Garden Street, Cambridge, MA, USA, 
         \email{tkaminsk@cfa.harvard.edu} \label{inst2}
    \and ESO, Alonso de Cordoba 3107, Vitacura, Santiago, Chile, \email{amehner@eso.org} \label{inst3}
          }
\abstract{CK\,Vul or Nova 1670 is an enigmatic eruptive object which underwent a stellar-merger event recorded by seventeenth-century observers. Its remnant was recently recovered at submillimeter wavelengths, revealing gas of an extraordinary isotopic composition indicative of past processing in the CNO cycles and partial He burning. Here, we present long-slit optical spectra of the remnant acquired with X-shooter at the Very Large Telescope at an unprecedented sensitivity and spectral coverage. The spectra cover features of key elements -- including H, He, C, N, and O -- at ionization degrees \small{I}--{\small III}. A classical analysis of the spectra allows us to identify several spatio-kinematic components in the probed part of the nebula at electron temperatures of 10--15\,kK and densities of 200--600\,cm$^{-3}$. We find that the nebula is most likely excited by shocks rather than by direct radiation of the stellar remnant. We provide a detailed analysis of the elemental abundances in the remnant and find that helium is twice more abundant than in the Sun. Nitrogen is also overabundant with a N/O ratio ten times larger than the solar one. These anomalous abundances strongly indicate that the observed gas was processed in CNO cycles of H burning, consistent with the submillimeter studies. Additionally, sub-solar abundances of heavier elements, such as Ne, S, and Ar, suggest that the progenitor of CK\,Vul was formed from material poorer in metals than the Sun and was therefore an old stellar system before the 1670 eruption.} 
\titlerunning{Elemental abundances in CK\,Vul}
\authorrunning{Tylenda, Kami\'nski, \& Mehner}   
\maketitle

\section{Introduction}\label{sec-intro}
CK Vulpeculae (CK\,Vul) is a remnant of a naked-eye stellar eruption observed in 1670--72 \citep{hevelius,shara85}. For a long time it was considered a classical nova (that is, a thermonuclear explosion on an accreting white dwarf), despite its unusual light curve that displayed three peaks over a time span of nearly three years. The stellar remnant is not seen in the optical but a faint hourglass nebula is visible around the presumable position of the 1670 nova \citep{shara82,shara85,hajduk2007}. The stellar object is however relatively bright in the far infrared \citep{evans2002,kamiNat}. This implies that the stellar remnant is obscured by an opaque dusty envelope whose sole presence is yet another argument against the classical nova scenario. Provoked by the enigma, \citet{kamiNat} launched a pioneering spectroscopic study of CK\,Vul at submillimeter (submm) wavelengths using the APEX telescope and the Submillimeter Array (SMA). The observations revealed that it is a strong and rich source of molecular emission. Kami\'nski et al. concluded that the 1670 eruption was that of a {\it red nova} rather than a classical nova -- a hypothesis already discussed in the past \citep{kato,tyl-blg360}.

It was the 2002 eruption of V838 Monocerotis \citep[e.g.,][]{munari} which raised a wide interest and allowed astrophysicists to define a new class of stellar eruptions named red novae or red transients. An increasingly red color implying a decreasing effective temperature is the primary characteristic of red novae. After the eruption, lasting a month up to a few years in the optical, the transients decline as very late M-type supergiants although they remain bright in the infrared for decades (or longer). Following a discussion of different mechanisms responsible for red-nova eruptions, \cite{ST2003} and next \cite{TS2006} proposed a binary merger as the scenario that best explains the observations of red novae. A 2008 eruption of V1309 Scorpii \citep{mason} was crucial for solving the red nova enigma. Using archival photometry of the object secured by the OGLE project \citep{udalski} and available for six years prior to the 2008 eruption, \citet{v1309} showed that before the eruption V1309\,Scorpii was a contact binary with a rapidly decreasing orbital period.

Submillimeter spectroscopy revealed a very rich molecular enviroment of CK\,Vul's remant \citep{kamiNat,kamiIRAM}. \cite{kamiIRAM} identified over 320 unique spectral features belonging to 27 molecules and their isotopologues. It has since been established that molecular emission at submm wavelengths is a common feature of Galactic red-nova remnants \citep{kamiSubmmNovae}. The great advantage of measuring rotational molecular lines at submm wavelengths is that they allow to determine the isotopic composition of the observed molecular matter. \cite{kamiIRAM} derived isotopic ratios of five elements in CK\,Vul: C, N, O, Si, and S. The isotopes of C, N, and O are of a particular importance as their isotopic ratios can reveal what nuclear burning processes occurred in the progenitor and, perhaps, during the merger. Kami\'nski et al. concluded that the gas currently seen in the remnant was processed in the CNO cycles of H-burning and underwent partial He burning.

Recently, very sensitive spectroscopy of CK\,Vul, acquired with mm-wave interferometers ALMA and NOEMA, led to a discovery of the rare isotopologue of aluminum fluoride that contains the radioactive isotope of aluminum, $^{26}$AlF \citep{kami26Al}. This was the first astronomical observation of a radioactive isotopologue in a stellar object, demonstrating once again the extraordinary properties of the CK\,Vul remnant. The observation yields a very low $^{27}$Al/$^{26}$Al ratio of about 7 which indicates that the gas in the remnant had to be recently processed in the hot Mg--Al cycles of H-burning. Most probably, the unstable isotope of Al comes from the outer layers of a He-core of a red giant disrupted in the merger that was witnessed in 1670--72. \citet{evans2018} questioned this interpretation and, based on a current low luminosity (1--10\,L$_{\odot}$) and potential overabundance of Li in the remnant \citep{hajduk2013}, proposed instead that Nova 1670 was a merger of a white dwarf and a brown dwarf.

Unfortunately, the submm spectroscopy alone cannot provide us with a quntitative measure of elemental abundances, which would be very useful for constraining the nature of the CK\,Vul's progenitor. Judging from a particularly high strength of lines of N-bearing molecules, \cite{kamiIRAM} suggested however that nitrogen is overabundant in the remnant \citep[see also][]{shara85}; similarly, the C/O ratio could be close to or slightly above one \citep[see also][]{evans}.

The optical nebula of CK\,Vul offers a possibility of deriving stringent constraints on the elemental abundances in the remnant. The abundances can be derived from nebular line intensities using well-established analysis methods commonly applied to planetary nebulae, \ion{H}{II} regions, and active galactic nuclei
\citep[e.g.][]{Osterbrock}. However, the nebular regions of CK\,Vul are very weak and spectra of CK\,Vul obtained thus far \citep{shara85,hajduk2007,hajduk2013} are of modest quality so that only a few strongest lines have been measured. This has hampered any advanced study of elemental abundances in this peculiar object. Here we present (Sects.\,\ref{sec-obs}--\ref{sec-spec}) and analyze (Sects.\,\ref{sec-spec}--\ref{sec-intens}) an optical spectrum of CK\,Vul nebula obtained at an unprecedented sensitivity. With this spectrum, we are able to derive the elemental abundances for the brightest part of the CK\,Vul nebula (Sect.\,\ref{sec-analyzis}) and discuss their origin (Sect.\,\ref{sec-discuss}).

\section{Observations}\label{sec-obs}
CK\,Vul was observed with X-shooter \citep{xshooter}  at the Very Large Telescope (VLT) on 29 Jul, and 15, 17, 18, and 27 Aug 2017. X-shooter records spectra simultaneously in three spectroscopic arms: UVB (300--560\,nm), VIS (550--1020\,nm), and NIR (1020--2480\,nm). The atmospheric dispersion corrector (ADC) was correcting the observations for the differential atmospheric refraction. This correction is particularly relevant for our observations as they took place at airmass 1.6--1.8. The visual seeing had a full-width at half-maximum (FWHM) of 0\farcs7--1\farcs3. The slit has a length of 11\arcsec. We set slit widths to 1\farcs3 for the UVB arm and to 1\farcs2 for the VIS and NIR arms. This resulted in spectral resolutions, $\lambda/\Delta\lambda$, of 4100, 6500, 4300 in the UVB, VIS, and NIR arms, respectively. The {\it stare} observing mode was used to increase the observing efficiency for the UVB and VIS arms but it compromised the quality of the NIR part of the spectra. With the large extent of the nebula of CK\,Vul of 71\arcsec\ \citep{hajduk2013}, it would be very difficult to find a nearby nodding position resulting in optimal sky extraction along the entire slit and therefore {\it stare} mode was a reasonable choice. The UVB and VIS detectors were used in their slow-readout mode with high gain and 1$\times$2 binning. The exposure times per observing block were 3030 and 3060\,s for UVB and VIS arms, respectively, adding to about 6\,h of on-source observations. Telluric standards, Hip095400, Hip095340, Hip090637, Hip107734, Hip101716, and Hip098609, and spectrophotometric standards, Feige\,110 and EG\,274, were observed with the same instrumental setup.

\begin{figure*}
\sidecaption
\includegraphics[angle=0, width=12cm]{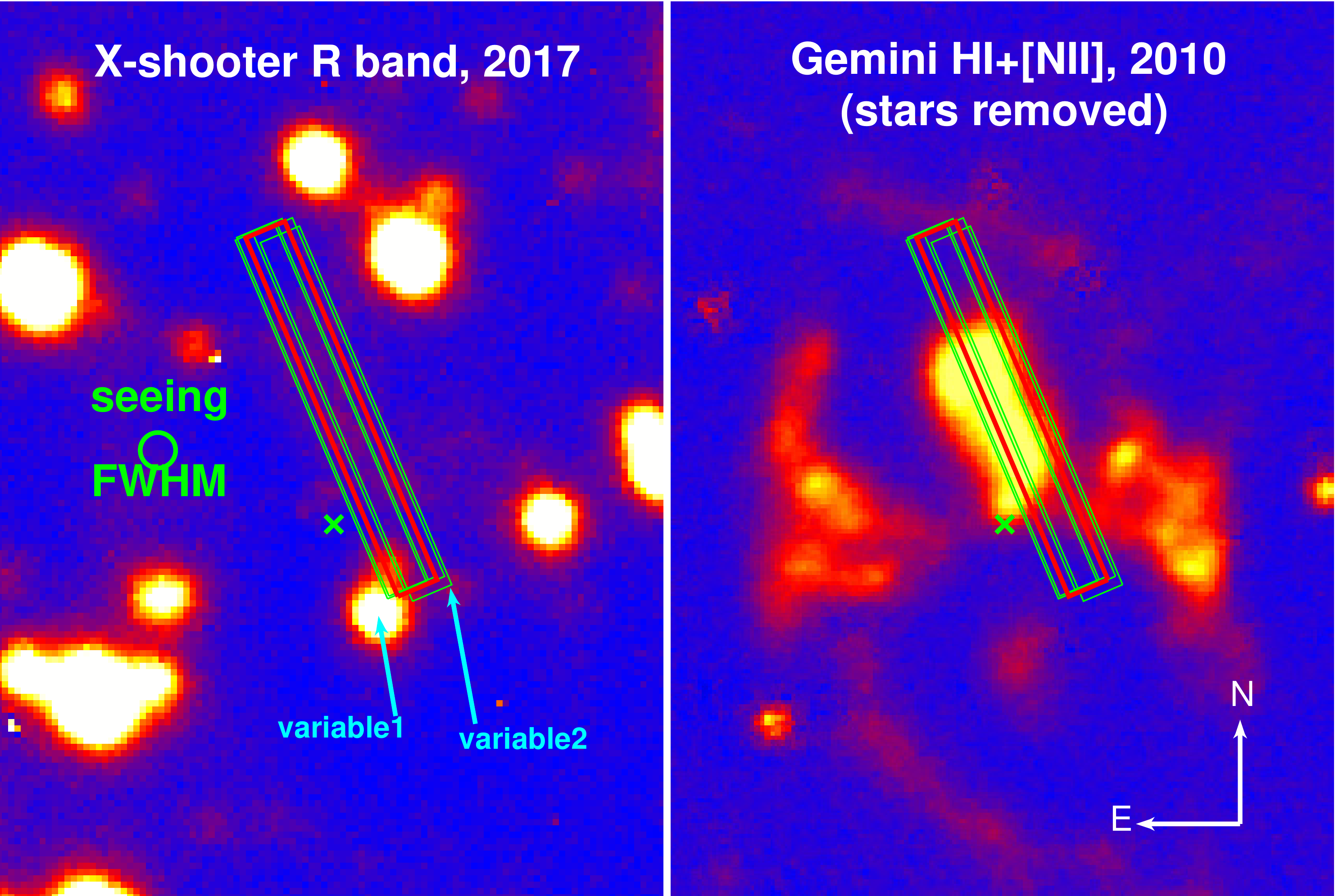}
\caption{Slit positions of X-shooter. The slit is represented by a box of a size of 11\arcsec$\times$1\farcs2 and at PA=13\degr. Positions corresponding to individual exposures are shown with green thin lines and their average position is shown with the red thick line. The background images are all in scale and coaligned.  Left: the average X-shooter acquisition image in $R$ band. Right: the image of the nebula in the H$\alpha$ filter from \citet{hajduk2013} with most point sources removed.}\label{fig-slit}
\end{figure*}

The observations were intended to secure a deep spectrum along the brightest part of the optical nebula. The nebula itself is too faint for direct pointing of the telescope. For this reason, the  observations were executed with the so-called {\it blind-offset} procedure in which the demanded position of the slit is given with respect to a brighter field star. This field source is first used for fine positioning of the telescope after which the telescope is slewed to the main target with the implemented offsets. The resulting slit positions are overlaid in Fig.\,\ref{fig-slit} on an average X-shooter acquisition image in the $R$-band filter and on the Gemini/GMOS image from \citet{hajduk2013} in the H$\alpha$ filter from 2010. Point sources were subtracted in the latter image to enhance the nebular emission. X-shooter acquisition images are taken immediately before each spectroscopic exposure. We corrected the images for bias and flat-field and combined them with {\it swarp} \citep{swarp}. We oriented the slit at a position angle (PA) of 13\fdg0 which is aligned with the northern jet of the atomic emission \citep[cf.][]{kamiNat}. The southern edge of the slit partially overlaps with two background stars. The brighter one \citet{hajduk2013} recognized as a variable source (their "variable 1"). The slit is close but does not overlap with the expected location of the stellar remnant of CK\,Vul. The predicted location of the star CK\,Vul, marked with a cross in Fig.\,\ref{fig-slit}, is based on observations of its submm continuum \citep{kami26Al}. The full extent of the slit corresponds to distances of 1\farcs4--8\farcs5 to North-East and 1\farcs4--3\farcs0 to South-West from that position.

The spectroscopic data were processed with the ESO X-shooter pipeline \citep{pipeline} version v3.1.0 with standard procedures resulting in two-dimensional spectra calibrated in wavelength and flux. The wavelength calibration has typical uncertainties of about 0.02, 0.03 nm in VIS and UVB spectra, respectively. Background sky correction turned out to be a very sensitive aspect of the data reduction because the nebula and field stars fill almost fully the slit and the off-source sky signal could not be adequately sampled. This problem particularly affected the removal of airglow emission lines. The VIS data were further corrected for telluric absorption with Molecfit version 1.4.0 \citep{molecfit1,molecfit2} using the telluric standards. This correction was imperfect, as well, and left residuals in the regions of the strongest telluric absorption. The wavelength scale of all spectra was then shifted to the heliocentric rest frame and spectra were combined. Some obvious instrumental artifacts were blanked in the spectra by hand. The two-dimensional spectra have a typical rms noise of 1.6$\times$10$^{-19}$ erg$\,$cm$^{-2}\,$s$^{-1}\,\AA^{-1}$ and the highest recorded signal-to-noise is of 665 at the default pipeline pixel binning of 0.2\,\AA$\times$0\farcs16.

\section{The spectrum and its spatial characteristics}\label{sec-spec}
\begin{figure*}
\centering
\includegraphics[angle=90, trim=100 100 180 100,width=0.65\textwidth]{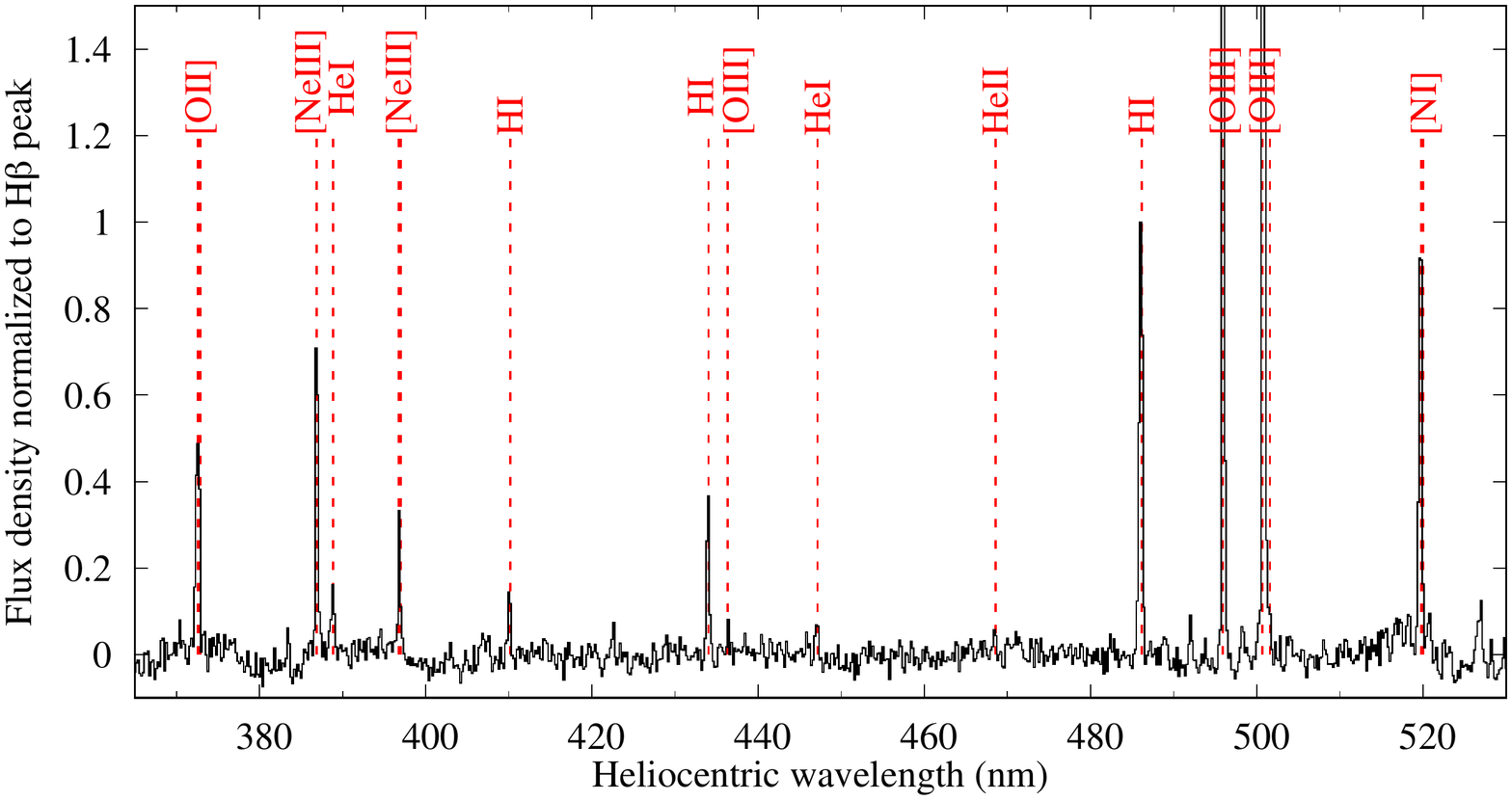}
\includegraphics[angle=90, trim=100 100 125 100,width=0.65\textwidth]{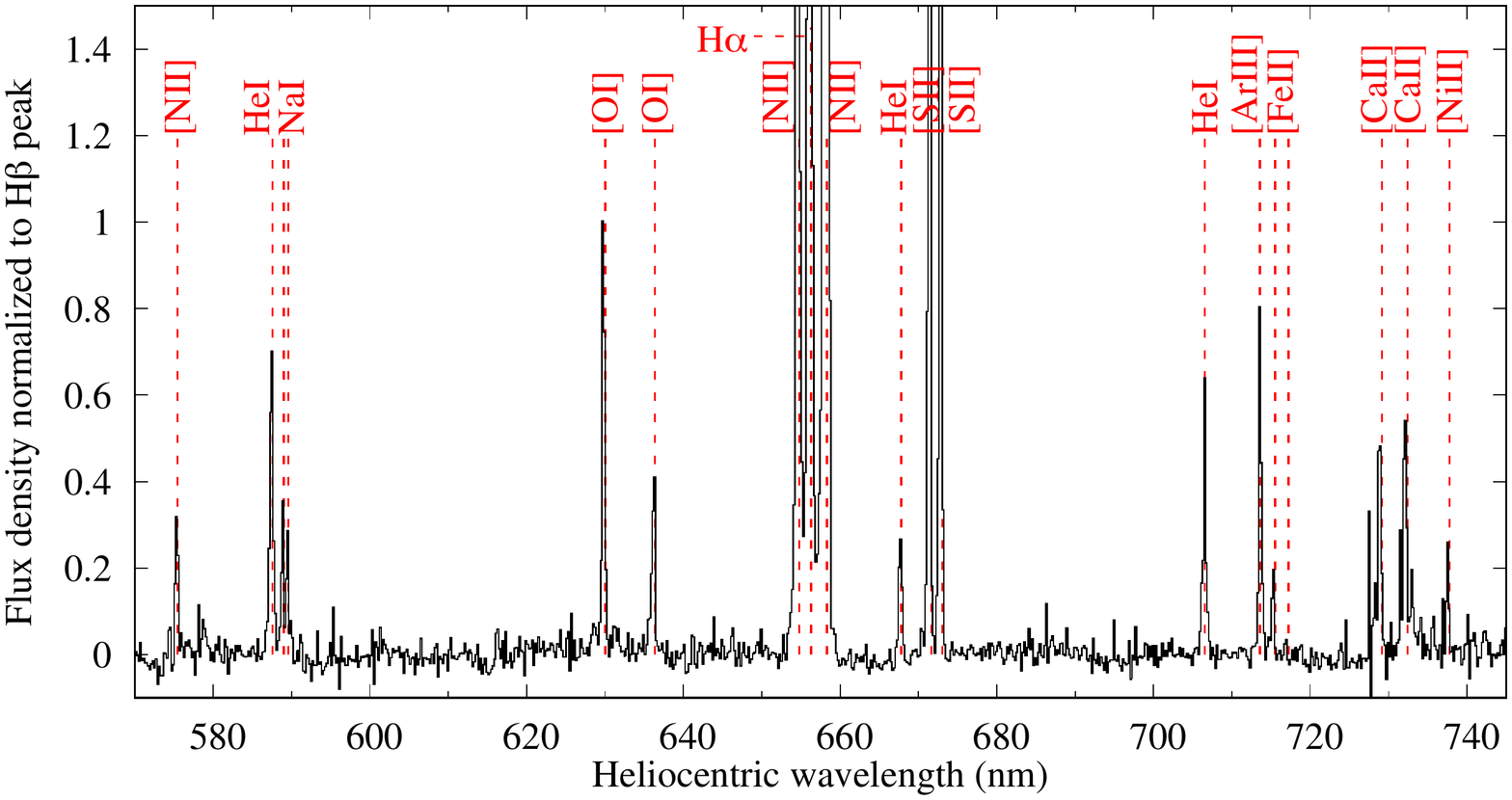}
\includegraphics[angle=90, trim=050 100 125 100,width=0.65\textwidth]{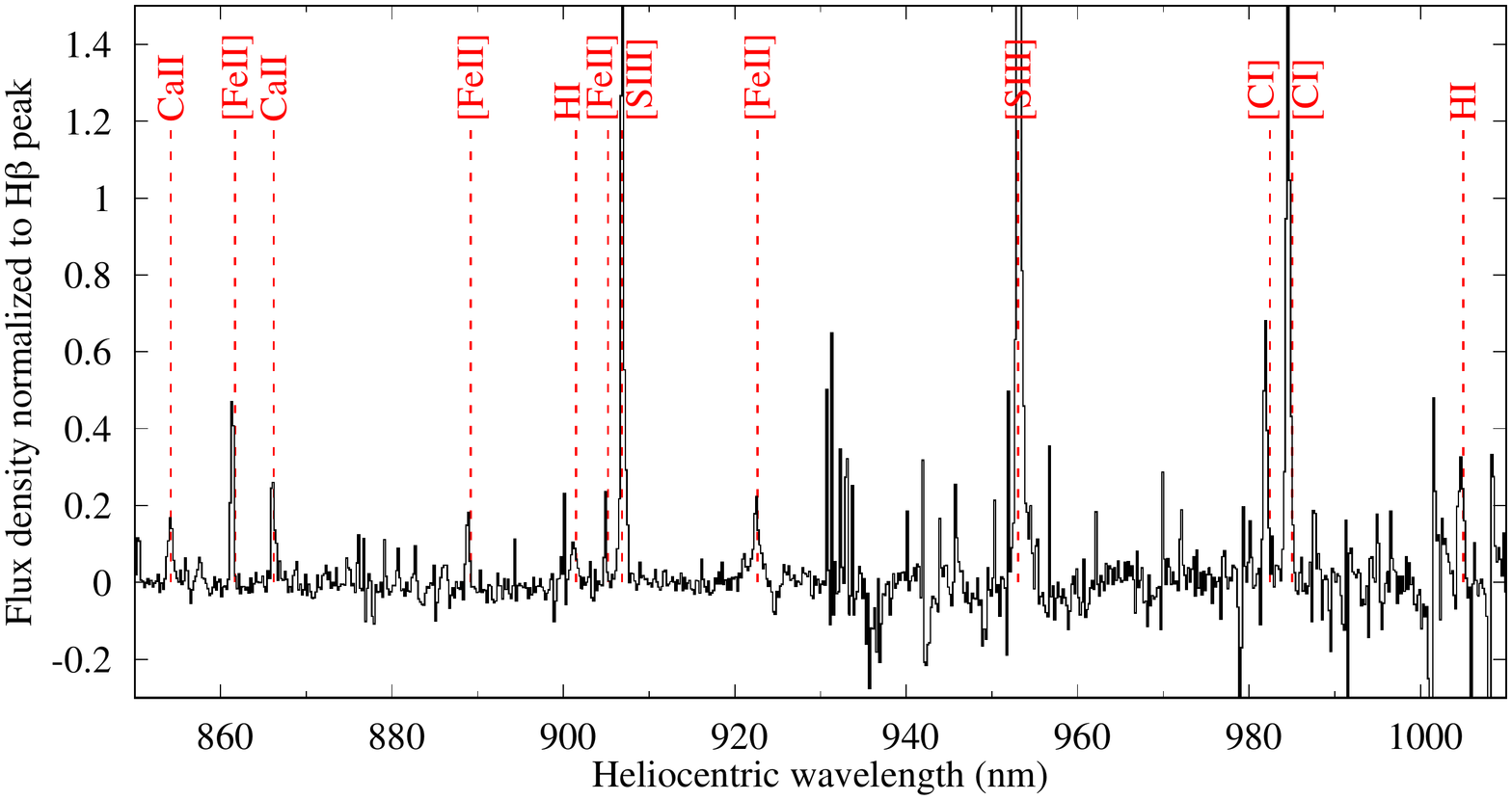}
\caption{Spectrum of the brightest nebular region of CK\,Vul obtained with X-shooter. The spectrum was obtained by summing the flux over the entire slit and smothing the spectral axis with 10 pixel average. Our identification of the major emission lines is shown in red. The vertical markers are drawn at rest laboratory wavelengths.}\label{fig-spec}
\end{figure*}

Figure\,\ref{fig-spec} presents a global spectrum obtained by summing the flux over the entire extent of the source along the slit. The upper panel shows the UBV spectrum while the middle and bottom panels present the VIS spectrum. We omit the spectrum between 740 and 850\,nm as it lacks any spectral features. The identification of the most relevant features is shown with red markers. The identified lines are also listed in Tables\,\ref{tab-fluxes-1}--\ref{tab-fluxes-3}. Note that at wavelengths longer than about 870\,nm the spectrum is often contaminated by residua of the imperfect telluric correction (see Sect.\,\ref{sec-obs}). Over 40 spectral features belonging to eleven elements (H, He, C, N, O, Ne, S, Ar, Fe, Ni, and Ca) at 
three ionization states (I--III) were identified. The presence of some lines, such as the $D$ doublet of \ion{Na}{I}, is doubtful owing to interstellar contamination.

Most lines typically observed in planetary nebulae can be found in the spectrum of CK\,Vul. However, lines of neutral species -- i.e., [\ion{O}{i}], [\ion{N}{i}], and [\ion{C}{i}] -- are significantly stronger in CK\,Vul than in a typical planetary nebula \citep[e.g.][]{gorny}. Also, emission of singly ionized metals -- i.e., [\ion{Fe}{ii}], [\ion{Ni}{ii}], and [\ion{Ca}{ii}] -- which are absent in the spectra of planetary nebulae, can easily be identified in Fig.\,\ref{fig-spec}. These differences may signify that the excitation mechanism of the CK\,Vul's nebular emission is different from that in planetary nebulae, i.e. is different than direct ionization by the radiation of the central star. We return to this point in Sect.\,\ref{sec-shocks}.

The quality of our spectrum warrants reliable estimates of the physical conditions in the probed part of the nebula, including the electron temperature, $T_{\rm e}$, and electron density, $N_{\rm e}$. The fluxes of lines [\ion{O}{iii}] 438.3, [\ion{N}{ii}] 575.5, and [\ion{S}{iii}] 631.2,\footnote{All lines in this paper are identified by their laboratory air wavelength in nm.} which are necessary to determine $T_{\rm e}$, can be reliably measured. Even the [\ion{C}{i}] 872.7 transition, although very weak, can be measured in one kinematical component, enabling another estimate of $T_{\rm e}$. The electron density can be derived from the line ratios of [\ion{O}{ii}] 372.6/372.9, [\ion{S}{ii}] 671.6/673.1, and [\ion{N}{i}] 519.8/520.0. The lines of [\ion{O}{ii}] and [\ion{N}{i}] are partially blended in our spectrum but are sufficiently well resolved to derive their ratios. Having determined $T_{\rm e}$ and $N_{\rm e}$, one can derive the ion abundances and then estimate the abundances of individual elements. For details of the nebular spectrum analysis, see e.g. Chapter 4 and 5 in \citet{Osterbrock}.

\begin{figure*}
\centering
\includegraphics[angle=0,trim=15 0 40 0, clip, width=0.485\textwidth]{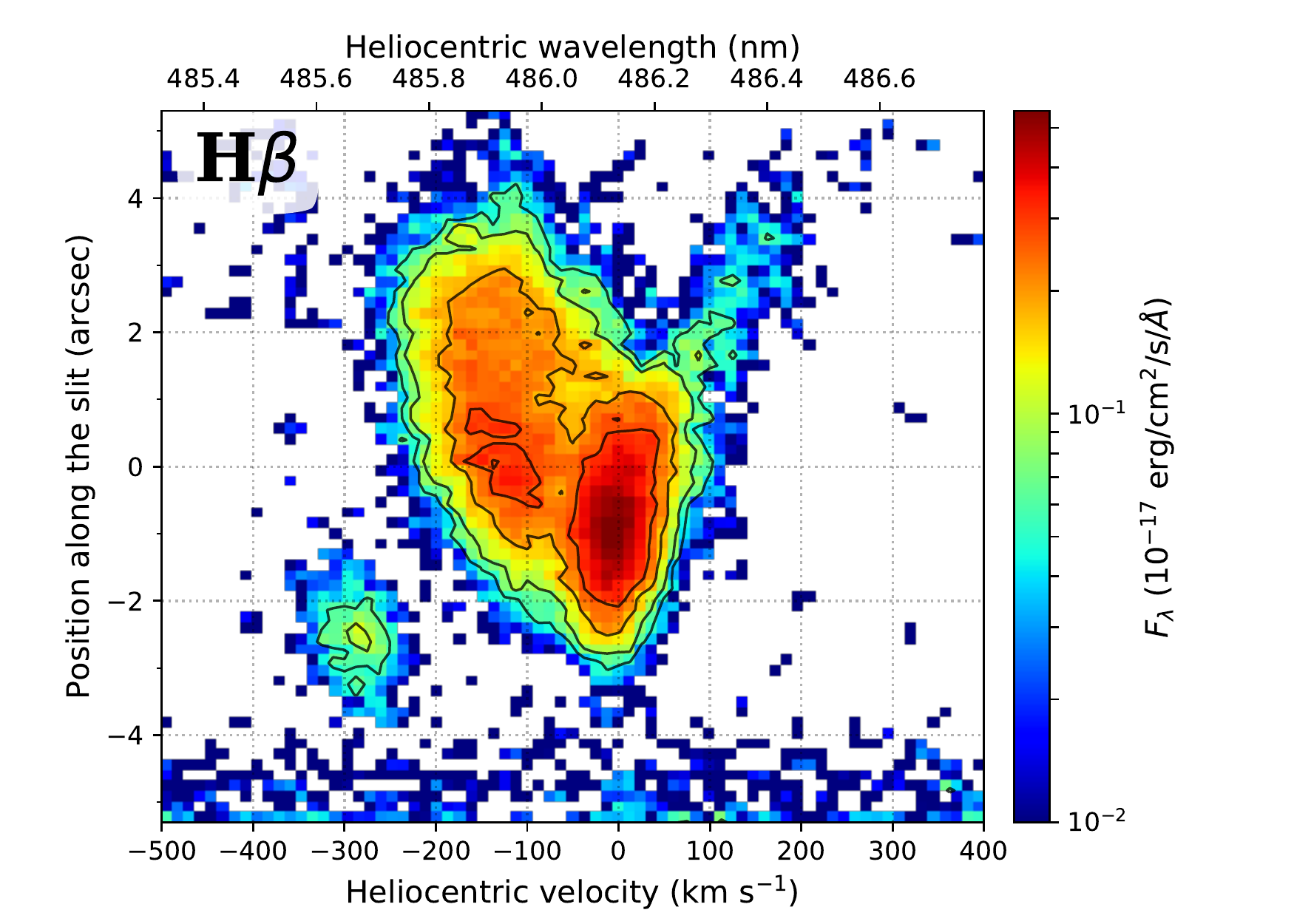}
\includegraphics[angle=0,trim=20 0 40 0, clip, width=0.48\textwidth]{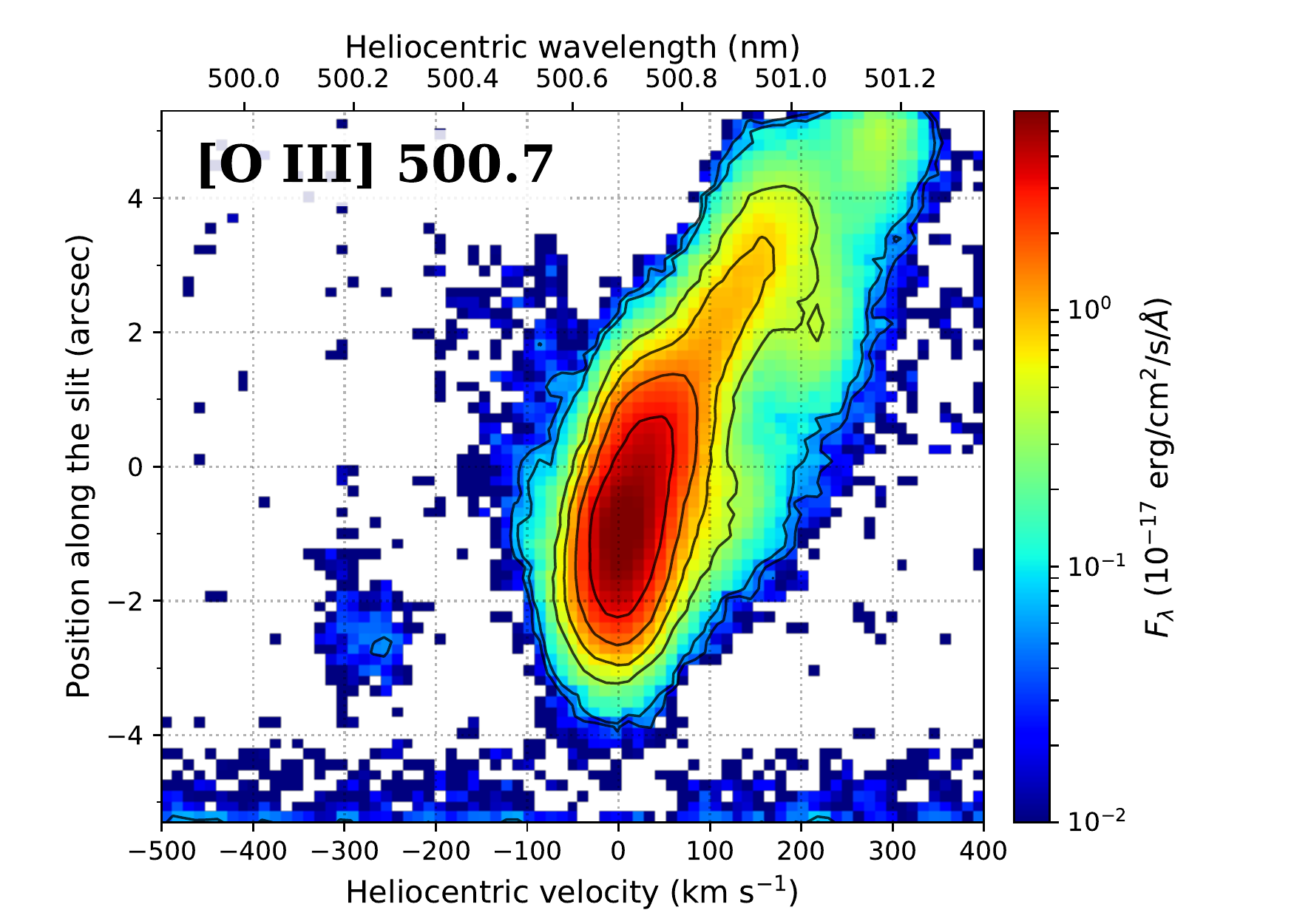}
\includegraphics[angle=0,trim=29 0 40 0, clip, width=0.46\textwidth]{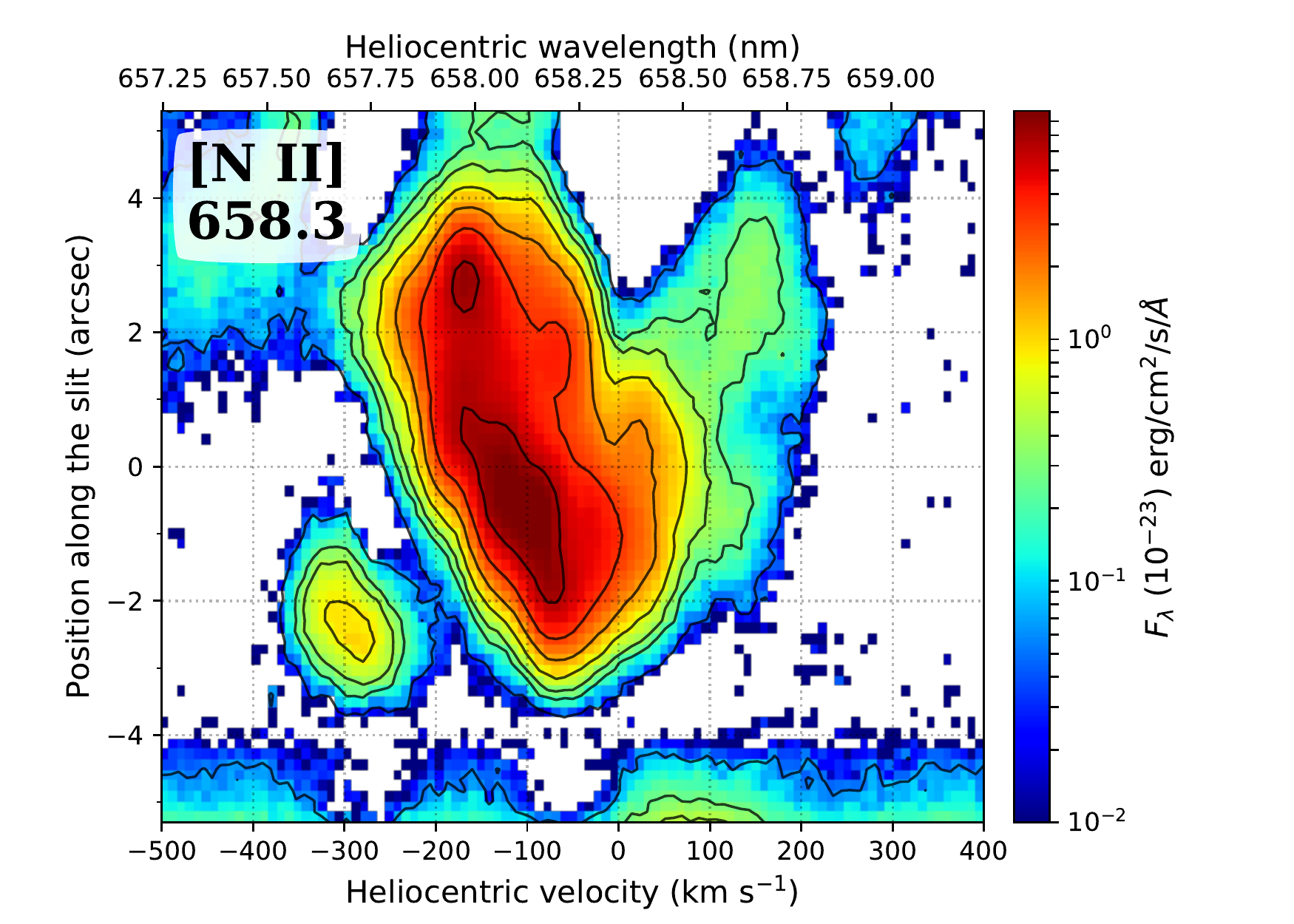}
\includegraphics[angle=0,trim=15 0 0 40, clip, width=0.42\textwidth]{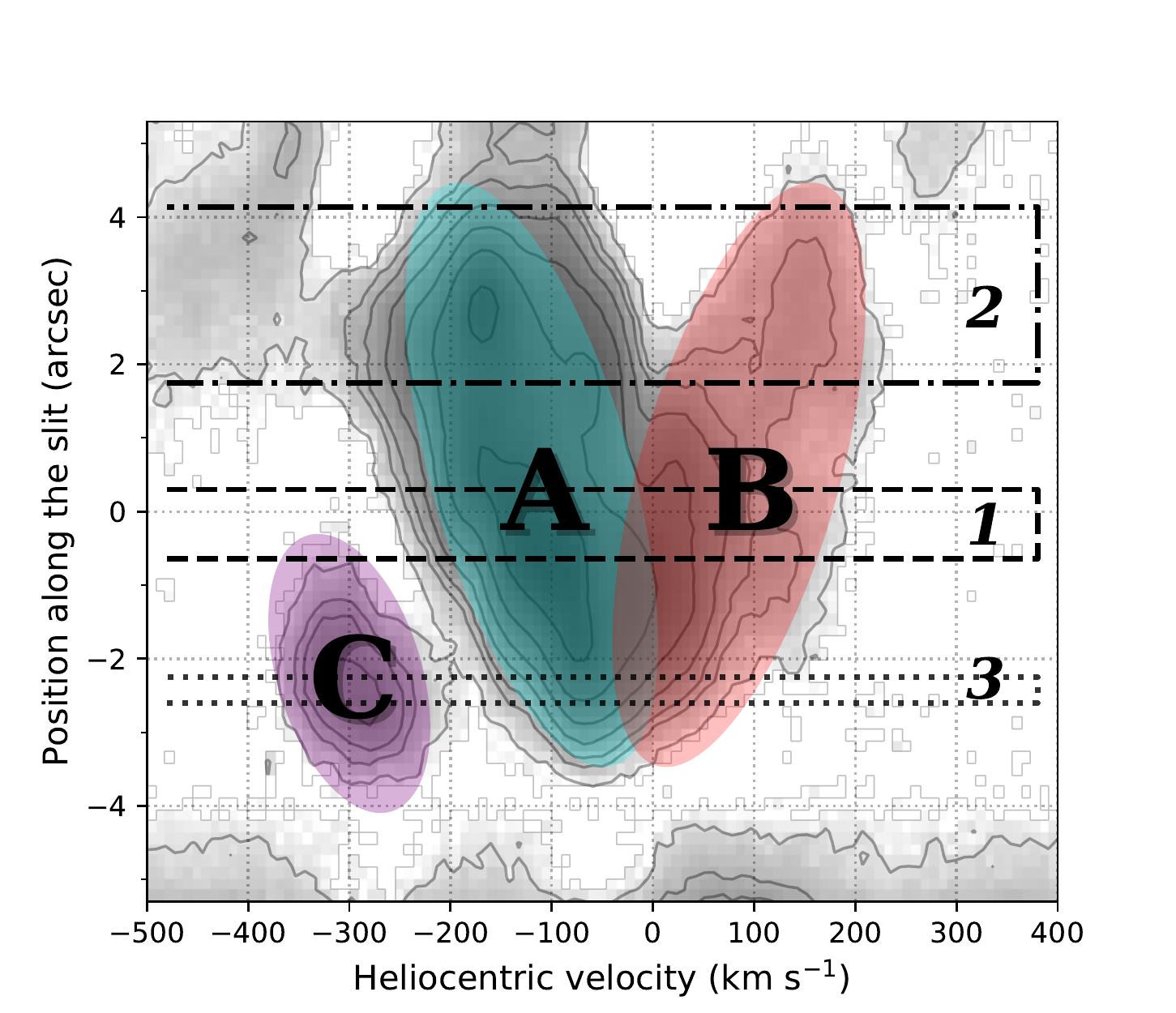}
\caption{Sample two-dimensional spectra of emission along the slit. Three emission lines are shown: H$\beta$ with contours at [3, 5, 10, 15]$\times\,\sigma$ noise levels; [\ion{O}{iii}] 500.7 with countours at [3, 5, 25, 50, 100, 200]$\times\,\sigma$; and [\ion{N}{ii}] 658.3 with contours at [3, 15, 25, 50, 100, 200, 400]$\times\,\sigma$. The lower right panel is the same map of [\ion{N}{ii}] (grey color scale) where we roughly deliminated with color shaded areas regions A, B, and C defined in the text. The dashed lines indicate the location of the three extracted spectra discussed in the text.}\label{fig-maps}
\end{figure*}

Profiles of the emission lines in CK\,Vul show a complex kinematical sub-structure, distinct for different species and ionization levels. This is illustrated in Fig.\,\ref{fig-maps} which shows two-dimensional distributions of emission in H$\beta$, [\ion{O}{iii}] 500.7, and [\ion{N}{ii}] 658.3 along the slit as a function of velocity (or wavelength). The emission covers a wide range of radial velocities, generally from $\sim$--350 to $\sim$+300\,\kms, but the velocity range depends on the line and the position along the slit. The main emission concentrates in two `wings' forming a structure with a shape of the letter `V' (Fig.\,\ref{fig-maps}). Hereafter, we refer to the left (negative-velocity) `wing' as component A and the right one as component B.  At a velocity of about --300\,\kms\ and a position $-2$\farcs5 along the slit, there is also a weaker emission feature which we identify as component C. The three components are roughly indicated in the bottom right panel of Fig.\,\ref{fig-maps}.

As can be seen in Fig.\,\ref{fig-maps}, the emission in H$\beta$ comes from all three components. This is obviously also the case for the other identified \ion{H}{i} lines. Emission of \ion{He}{i} is observed in components A and B but is almost absent in component C. Instead, the [\ion{N}{ii}] emission is concentrated in components A and C. This is also true for all lines of singly ionized and neutral species, i.e. [\ion{O}{i}], [\ion{O}{ii}], [\ion{N}{i}], [\ion{S}{ii}], etc. Component B dominates the emission in the [\ion{O}{iii}] lines and the lines from other doubly ionized species ([\ion{Ne}{iii}], [\ion{S}{iii}], and [\ion{Ar}{iii}]). The components readily differ in  excitation conditions. Specifically, the excitation of plasma in component B is significantly higher than in components A and C.

Because of these excitation differences, we did not analyze the global spectrum presented in Fig.\,\ref{fig-spec}, but rather disentangled the spectra of individual spatio-kinematic components and analyzed them separately. However, A and B are partially blended and the degree of blending depends on the range of pixels chosen for summing along the slit. After several attempts to optimize this separation, we took three spatial cuts shown in the bottom right panel of Fig.\,\ref{fig-maps} with dashed lines. {\it Spectrum 1} was obtained by summing emission in pixels located within a region between $l$=--0\farcs64 to 0\farcs30 with respect to the slit center (with negative values representing the direction towards South-West). This central spectrum extraction includes the maximum emission of the strongest lines. {\it Spectrum 2} was extracted by adding fluxes for 1\farcs75$\leq l \leq\,$4\farcs14. In this spectrum, components A and B are better separable than in {\it spectrum 1} but the line emission is weaker, especially in component B. A third spectrum was produced by summing the flux in  
--2\farcs60$\leq l \leq\,$--2\farcs25, meant to study component C.

\section{Line intensities and extinction}\label{sec-intens}
\begin{figure}
\centering
\includegraphics[angle=0,trim=0 0 0 0, clip, width=0.5\textwidth]{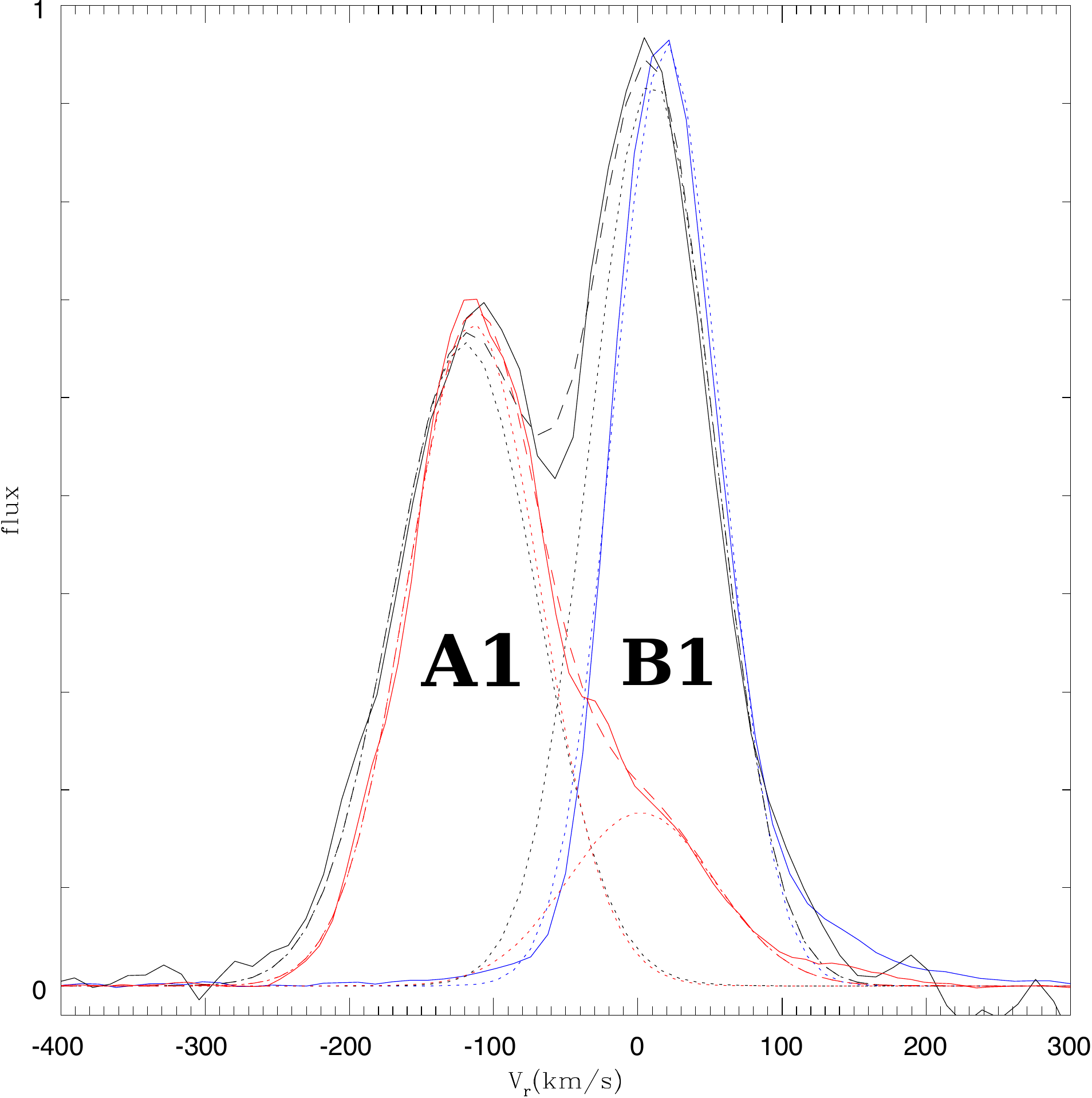}
\caption{Emission line profiles in spectrum 1. Full curves represent observed profiles of H$\beta$ (black), [\ion{N}{ii}] 658.3 (red), and [\ion{O}{iii}] 500.7 (blue). Dotted curves show Gaussian components fitted to the observed profiles. The dashed curve is the sum of the Gaussian components.}\label{fig-prof1}
\end{figure}

Figure\,\ref{fig-prof1} displays typical line profiles in spectrum 1, representing the central extraction. The H$\beta$ emission displays two clear peaks corresponding to components A and B. The [\ion{O}{iii}] line shows a single component arising from region B. Contrary to this, the [\ion{N}{ii}] line arises  mainly in component A.

\begin{figure}
\centering
\includegraphics[angle=0,trim=0 0 0 0, clip, width=0.5\textwidth]{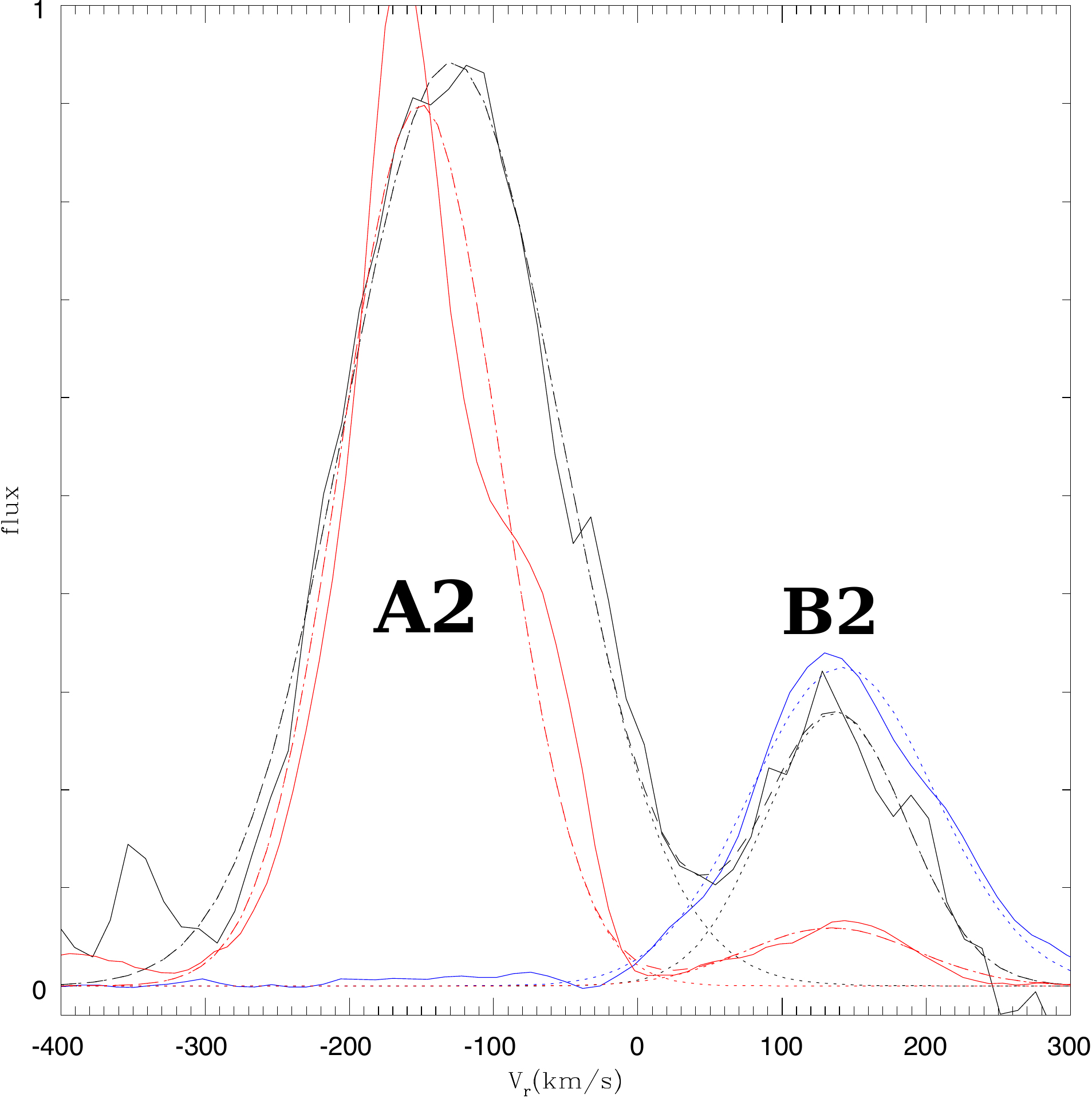}
\caption{The same as Fig.\,\ref{fig-prof1} but for spectrum 2.}\label{fig-prof2}
\end{figure}

As is best seen in the H$\beta$ profile in Fig.\,\ref{fig-prof1}, the emission of components A and B is partially blended. We disentangled them by fitting two Gaussian profiles to the observed features. The individual Gaussian profiles are plotted with dotted curves and their sum is shown with a dashed curve. We adopted the integral over the fitted Gaussian as the flux of the corresponding component. To have consistent measurements, the procedure of fitting a Gaussian profile was also applied to lines showing a single component, e.g to the [\ion{O}{iii}] line in Fig.\,\ref{fig-prof1}. 

The results of the Gaussian fits to the emission line profiles of spectrum 1 are presented in Table\,\ref{tab-fluxes-1}. They include measurements of the flux (within the defined extraction aperture which was different for the three spectra), relative flux (scaled so that H$\beta$ has a value of 100), central radial velocity ($V_r$), FWHM, and an estimate of flux uncertainties (column headed "Note"). 

\begin{table*}[]\centering \small
\caption{Measurements of emission lines in spectrum 1.}\label{tab-fluxes-1}
\begin{tabular}{lc | rrccc | rrccc}
\hline
   &    & \multicolumn{5}{c|}{component A}&\multicolumn{5}{c}{component B}\\
\multicolumn{1}{c}{$\lambda_0$} & \multicolumn{1}{c|}{Ion} &  
\multicolumn{1}{c}{Flux}& \multicolumn{1}{c}{Relative} & $V_r$ & FWHM & Note & 
\multicolumn{1}{c}{Flux}& \multicolumn{1}{c}{Relative} & $V_r$ & FWHM & Note \\
\multicolumn{1}{c}{(nm)}& \multicolumn{1}{c|}{}&
\multicolumn{1}{c}{(10$^{-16}$\,erg}&\multicolumn{1}{c}{flux} & (\kms) & (\kms)& &
\multicolumn{1}{c}{(10$^{-16}$\,erg}& \multicolumn{1}{c}{flux} & (\kms) & (\kms)& \\
& \multicolumn{1}{c|}{}&
\multicolumn{1}{c}{cm$^{-2}$\,s$^{-1}$)}&\multicolumn{1}{c}{(H$\beta$=100)}&&&&
\multicolumn{1}{c}{cm$^{-2}$\,s$^{-1}$)}&\multicolumn{1}{c}{(H$\beta$=100)}&&&\\
\hline\hline
~~372.60& {[}\ion{O}{ii}]   &0.200&50.0&   --90 &132& a  &          &&      &      &     \\ 
~~372.88& {[}\ion{O}{ii}]   &0.184&46.0&  --100 &120& a  &          &&      &      &     \\ 
~~386.88& {[}\ion{Ne}{iii}] &        &&       &      &    &0.574&118.6&16&89& a   \\ 
~~388.86&  \ion{He}{i}    &0.057&14.3&  --108 &94& c  &0.069&14.3&24&113& c   \\ 
~~396.75& {[}\ion{Ne}{iii}] &        &&       &      &    &0.145&30.0&18&94& a   \\ 
~~397.01&  \ion{H}{i}     &0.022&5.5&  --120 &130& c  &0.027&5.6&10&101& c   \\ 
~~410.17&  \ion{H}{i}     &0.056&14.0&  --120 &144& b  &0.064&13.2&10&101& b   \\ 
~~434.05&  \ion{H}{i}     &0.131&32.8&  --120 &118& b  &0.152&31.4&10&99& b   \\ 
~~436.32& {[}\ion{O}{iii}]  &        &&       &      &    &0.054&11.2&17&78& b   \\ 
~~447.15&  \ion{He}{i}    &0.028&7.0&  --102 &104& c  &0.042&8.7&11&87& c   \\ 
~~468.57&  \ion{He}{ii}   &0.024&6.0&  --118 &89& c  &0.030&6.2&22&92& c   \\ 
~~486.13&  \ion{H}{i}     &0.400&100.0&  --120 &118& a  &0.484&100.0&10&101& a   \\ 
~~495.89& {[}\ion{O}{iii}]  &        &&       &      &    &1.632&337.2&19&87& a   \\ 
~~500.68& {[}\ion{O}{iii}]  &        &&       &      &    &5.064&1046.3&20&87& a   \\ 
~~501.57&  \ion{He}{i}    &0.034&8.5&  --107 &101& c  &0.046&9.5&11&92& c   \\ 
~~519.79& {[}\ion{N}{i}]    &0.280&70.0&  --123 &118& b  &          &&      &      &     \\ 
~~520.03& {[}\ion{N}{i}]    &0.340&85.0&  --125 &120& b  &          &&      &      &     \\ 
~~575.46& {[}\ion{N}{ii}]   &0.123&30.8&  --103 &106& b  &0.160&33.1&9&89& b   \\ 
~~587.56&  \ion{He}{i}    &0.250&62.5&  --129 &97& a  &0.403&83.3&8&106& a   \\ 
~~630.03& {[}\ion{O}{i}]    &0.480&120.0&  --116 &101& b  &          &&      &      &     \\ 
~~631.21& {[}\ion{S}{iii}]  &        &&       &      &    &0.060&12.4&17&87& b   \\ 
~~636.38& {[}\ion{O}{i}]    &0.175&43.8&  --118 &115& b  &          &&      &      &     \\ 
~~654.80& {[}\ion{N}{ii}]   &4.296&1074.0&  --114 &106& a  &1.288&266.1&2&120& b   \\ 
~~656.28&  \ion{H}{i}     &3.240&810.0&  --138 &92& a  &4.000&826.4&   --2 &118& a   \\ 
~~658.34& {[}\ion{N}{ii}]   &13.500&3375.0&  --114 &108& a  &3.920&809.9&2&120& b   \\ 
~~667.82&  \ion{He}{i}    &0.116&29.0&  --123 &101& b  &0.145&30.0&12&94& b   \\ 
~~671.64& {[}\ion{S}{ii}]   &1.648&412.0&  --125 &97& a  &          &&      &      &     \\ 
~~673.08& {[}\ion{S}{ii}]   &1.600&400.0&  --125 &97& a  &          &&      &      &     \\ 
~~706.52&  \ion{He}{i}    &0.082&20.5&  --136 &80& b  &0.413&85.3&17&101& a   \\ 
~~713.58& {[}\ion{Ar}{iii}] &        &&       &      &    &0.547&113.0&13&99& a   \\ 
~~715.52& {[}\ion{Fe}{ii}]  &0.145&36.3&  --104 &108& b  &          &&      &      &     \\ 
~~717.20& {[}\ion{Fe}{ii}]  &0.024&6.0&  --113 &108& c  &          &&      &      &     \\ 
~~729.15& {[}\ion{Ca}{ii}]  &0.419&104.8&  --112 &141& b  &          &&      &      &     \\ 
~~732.39& {[}\ion{Ca}{ii}]  &0.445&111.3&  --123 &146& b  &          &&      &      &     \\ 
~~737.78& {[}\ion{Ni}{ii}]  &0.205&51.3&  --103 &115& b  &          &&      &      &     \\ 
~~745.26& {[}\ion{Fe}{ii}]  &0.052&13.0&  --102 &92& c  &          &&      &      &     \\ 
~~861.70& {[}\ion{Fe}{ii}]  &0.461&115.3&  --103 &115& b  &          &&      &      &     \\ 
~~901.49&  \ion{H}{i}     &0.048&12.0&  --138 &80& c  &0.052&10.7&   --2 &118& c   \\ 
~~905.20& {[}\ion{Fe}{ii}]  &0.138&34.5&  --100 &92& b  &          &&      &      &     \\ 
~~906.86& {[}\ion{S}{iii}]  &0.075&18.8&  --110 &54& c  &1.280&264.5&25&115& a   \\ 
~~922.66& {[}\ion{Fe}{ii}]  &0.110&27.5&   --95 &141& b  &          &&      &      &     \\ 
~~922.90&  \ion{H}{i}     &0.067&16.8&  --135 &82& c  &0.083&17.1&22&134& c   \\ 
~~953.06& {[}\ion{S}{iii}]  &0.196&49.0&  --110 &54& b  &3.620&747.9&25&115& a   \\ 
~~954.60&  \ion{H}{i}     &0.079&19.8&  --134 &80& c  &0.096&19.8&   --6 &82& c   \\ 
~~982.41& {[}\ion{C}{i}]    &0.372&93.0&  --128 &141& b  &          &&      &      &     \\ 
~~985.03& {[}\ion{C}{i}]    &0.946&236.5&  --130 &127& a  &          &&      &      &     \\ 
1004.98&  \ion{H}{i}     &0.183&45.8&  --147 &99& c  &0.225&46.5&  --14 &122& c   \\ 
\hline\end{tabular}
\tablefoot{The first two columns identify the lines including their laboratory 
wavelength, $\lambda_0$. The five subsequent columns present results of 
Gaussian fits to component A and the last five columns to component B. The fluxes are given in absolute units and relative to H$\beta$ ($\times$100). They represent the same region along the spatial axis for regions A and B but different than that used in Tables\,\ref{tab-fluxes-2} and \ref{tab-fluxes-3}. 
$V_r$ is the central radial velocity of the line derived in the fit. Last column quantifies the flux measurements: a, flux accurate to within 10\%; b, error between 10 and 30\%; c, error greater than 30\%.} 
\end{table*}

\begin{table*}[]\centering \small
\caption{The same as Table\,\ref{tab-fluxes-1} but for spectrum 2.}\label{tab-fluxes-2}
\begin{tabular}{lc | rrccc | rrccc}
\hline
   &    & \multicolumn{5}{c|}{component A}&\multicolumn{5}{c}{component B}\\
\multicolumn{1}{c}{$\lambda_0$} & \multicolumn{1}{c|}{Ion} &  
\multicolumn{1}{c}{Flux}& \multicolumn{1}{c}{Relative} & $V_r$ & FWHM & Note & 
\multicolumn{1}{c}{Flux}& \multicolumn{1}{c}{Relative} & $V_r$ & FWHM & Note \\
\multicolumn{1}{c}{(nm)}& \multicolumn{1}{c|}{}&
\multicolumn{1}{c}{(10$^{-16}$\,erg}&\multicolumn{1}{c}{flux} & (\kms) & (\kms)& &
\multicolumn{1}{c}{(10$^{-16}$\,erg}& \multicolumn{1}{c}{flux} & (\kms) & (\kms)& \\
& \multicolumn{1}{c|}{}&
\multicolumn{1}{c}{cm$^{-2}$\,s$^{-1}$)}&\multicolumn{1}{c}{(H$\beta$=100)}&&&&
\multicolumn{1}{c}{cm$^{-2}$\,s$^{-1}$)}&\multicolumn{1}{c}{(H$\beta$=100)}&&&\\
\hline\hline
372.60& {[}\ion{O}{ii}]   &0.320&37.1&  --124 &167& a &           &&       &       &  \\
372.88& {[}\ion{O}{ii}]   &0.326&37.8&  --124 &167& a &           &&       &       &  \\
386.88& {[}\ion{Ne}{iii}] &0.063&7.3&  --126 &203& c &0.358&214.4&140&148& a\\
388.86&  \ion{He}{i}    &0.109&12.6&  --119 &134& b &0.044&26.3&118&120& c\\
396.75& {[}\ion{Ne}{iii}] &0.021&2.4&  --126 &188& d &0.105&62.9&124&92& b\\
410.17&  \ion{H}{i}     &0.131&15.2&  --118 &177& b &0.015&9.0&137&118& d\\
434.05&  \ion{H}{i}     &0.292&33.9&  --122 &179& b &0.053&31.7&137&118& b\\
436.32& {[}\ion{O}{iii}]  &          &&       &      &   &0.020&12.0&145&148& d\\
447.15&  \ion{He}{i}    &0.075&8.7&  --125 &193& c &           &&       &       &  \\
486.13&  \ion{H}{i}     &0.862&100.0&  --129 &177& a &0.167&100.0&137&115& a\\
495.89& {[}\ion{O}{iii}]  &0.023&2.7&  --127 &106& d &0.960&574.9&144&153& a\\
500.68& {[}\ion{O}{iii}]  &0.069&8.0&  --127 &106& c &2.960&1772.5&142&151& a\\
519.79& {[}\ion{N}{i}]    &0.800&92.8&  --146 &94& a &           &&       &       &  \\
520.03& {[}\ion{N}{i}]    &0.869&100.8&  --146 &94& a &           &&       &       &  \\
575.46& {[}\ion{N}{ii}]   &0.188&21.8&  --137 &127& c &           &&       &       &  \\
587.56&  \ion{He}{i}    &0.543&63.0&  --147 &160& a &0.176&105.4&141&137& b\\
630.03& {[}\ion{O}{i}]    &0.937&108.7&  --157 &92& a &           &&       &       &  \\
636.38& {[}\ion{O}{i}]    &0.338&39.2&  --159 &99& b &           &&       &       &  \\
654.80& {[}\ion{N}{ii}]   &6.752&783.3&  --151 &127& a &0.716&428.7&134&139& b\\
656.28&  \ion{H}{i}     &7.240&839.9&  --149 &146& a &1.352&809.6&136&134& a\\
658.34& {[}\ion{N}{ii}]   &21.500&2494.2&  --151 &130& a &1.512&905.4&134&139& a\\
667.82&  \ion{He}{i}    &0.250&29.0&  --128 &177& b &0.072&43.1&107&174& c\\
671.64& {[}\ion{S}{ii}]   &3.526&409.0&  --153 &104& a &           &&       &       &  \\
673.08& {[}\ion{S}{ii}]   &3.276&380.0&  --154 &104& a &           &&       &       &  \\
706.52&  \ion{He}{i}    &0.187&21.7&  --160 &134& b &0.197&118.0&154&141& b\\
713.58& {[}\ion{Ar}{iii}] &          &&       &      &   &0.247&147.9&145&127& a\\
715.52& {[}\ion{Fe}{ii}]  &0.159&18.4&  --134 &151& b &           &&       &       &  \\
729.15& {[}\ion{Ca}{ii}]  &0.640&74.2&  --140 &137& b &           &&       &       &  \\
732.39& {[}\ion{Ca}{ii}]  &0.442&51.3&  --141 &148& b &           &&       &       &  \\
737.78& {[}\ion{Ni}{ii}]  &0.203&23.5&  --132 &139& b &           &&       &       &  \\
861.70& {[}\ion{Fe}{ii}]  &0.530&61.5&  --138 &148& b &           &&       &       &  \\
872.71& {[}\ion{C}{i}]    &0.080&9.3&  --160 &106& c &           &&       &       &  \\
889.19& {[}\ion{Fe}{ii}]  &0.227&26.3&  --134 &165& b &           &&       &       &  \\
901.49&  \ion{H}{i}     &0.113&13.1&  --136 &118& c &           &&       &       &  \\
906.86& {[}\ion{S}{iii}]  &0.100&11.6&  --161 &179& d &0.473&283.2&164&111& a\\
922.66& {[}\ion{Fe}{ii}]  &0.074&8.6&  --160 &94& c &           &&       &       &  \\
922.90&  \ion{H}{i}     &0.143&16.6&  --129 &80& c &           &&       &       &  \\
953.06& {[}\ion{S}{iii}]  &0.328&38.1&  --161 &179& c &1.568&938.9&166&144& a\\
982.41& {[}\ion{C}{i}]    &1.344&155.9&  --161 &104& b &           &&       &       &  \\
985.03& {[}\ion{C}{i}]    &3.816&442.7&  --157 &97& a &           &&       &       &  \\
\hline\end{tabular}
\tablefoot{The first two columns identify the lines including their laboratory 
wavelength, $\lambda_0$. The five subsequent columns present results of 
Gaussian fits to component A and the last five columns to component B. The fluxes are given in absolute units and relative to H$\beta$ ($\times$100). They represent the same region along the spatial axis for regions A and B but different than that used in Tables\,\ref{tab-fluxes-1} and \ref{tab-fluxes-3}. 	 
$V_r$ is the central radial velocity of the line. Last column quantifies the flux measurements: a, flux accurate to within 10\%; b, error between 10 and 30\%; c, error greater than 30\%; d, line marginally detected.} 
\end{table*}

The radial velocity of component A in spectrum 1, calculated as the mean radial velocity of lines with good Gaussian fits, is $\langle V_r\rangle$=--116\,\kms. The mean FWHM, calculated in the same way is $\langle\theta\rangle$=112\,\kms. This value includes the instrumental broadening. FWHM of the instrumental profile, as estimated from airglow lines, is $\simeq$ 35\,\kms. Thus, FWHM of component A in spectrum 1 corrected for instrumental broadening is $\langle\theta_c\rangle$=106\,\kms. The same figures for component B are 13\,\kms\ and 100\,\kms\ ($\langle\theta_c\rangle$=94\,\kms), respectively.

Figure\,\ref{fig-prof2} is analogues to Fig.\,\ref{fig-prof1} and displays line profiles and Gaussian fits for spectrum 2 extracted in the East-North part of the slit. In this region, components A and B are easier to separate than in spectrum 1. Component B is much weaker than component A. The profile of [\ion{N}{ii}] 658.3 in component A is not well reproduced by the Gaussian. It is narrower and shifted to more negative velocities than the profile of H$\beta$. Readily, component A has a somewhat more complex structure in spectrum 2 than in 1. The derived line parameters are summarized in Table\,\ref{tab-fluxes-2}. The meaning of the columns is the same  as in Table\,\ref{tab-fluxes-1}.

In spectrum 2, the mean radial velocity and the mean FWHM for component A are $\langle V_r\rangle$=--141\,\kms\ and $\langle\theta\rangle$=136\,\kms\ ($\langle\theta_c\rangle$=131\,\kms), while for component B these figures are $\langle V_r\rangle$=143\,\kms\ and $\langle\theta\rangle$=132\,\kms\ ($\langle\theta_c\rangle$=127\,\kms). 
The separation of the components in spectrum 2 is greater than in spectrum 1 and the line profiles in spectrum 1 
are somewhat wider. The latter is partially due to the larger range of pixels included in extraction of spectrum 2 compared to spectrum 1.

Spectrum 3 was obtained solely to study component C (see Fig.\,\ref{fig-maps}). There is a component produced by blending A and B in this spectrum but it is well separated from C and does not affect our fits. The measurements are presented in Table\,\ref{tab-fluxes-3}. The mean radial velocity of component C is $\langle V_r\rangle$=--285\,\kms, 
while the mean FWHM of the lines is $\langle\theta\rangle$=84\,\kms ($\langle\theta_c\rangle$=76\,\kms).

\begin{table}[]\centering \small
\caption{The same as Table\,\ref{tab-fluxes-1} but for component C in spectrum 3.}\label{tab-fluxes-3}
\begin{tabular}{lc | rrccc}
\hline
\multicolumn{1}{c}{$\lambda_0$} & \multicolumn{1}{c|}{Ion} &  \multicolumn{1}{c}{Flux}& \multicolumn{1}{c}{Relative} & $V_r$ & FWHM & Note \\
\multicolumn{1}{c}{(nm)}& \multicolumn{1}{c|}{}&\multicolumn{1}{c}{(10$^{-16}$\,erg}& \multicolumn{1}{c}{flux} & (km & (\kms)& \\
& \multicolumn{1}{c|}{}&\multicolumn{1}{c}{cm$^{-2}$\,s$^{-1}$)}&\multicolumn{1}{c}{(H$\beta$=100)}&\multicolumn{1}{c}{s$^{-1}$)}&&\\
\hline\hline
372.60& {[}\ion{O}{ii}]  &0.0292&81.8& --279 &85&b \\
372.88& {[}\ion{O}{ii}]  &0.0310&86.8& --273 &97&b \\
434.05&  \ion{H}{i}    &0.0131&36.7& --278 &82&b \\
486.13&  \ion{H}{i}    &0.0357&100.0& --282 &82&a \\
495.89& {[}\ion{O}{iii}] &0.0082&23.0& --274 &104&c \\
500.68& {[}\ion{O}{iii}] &0.0241&67.5& --276 &99&b \\
519.79& {[}\ion{N}{i}]   &0.0136&38.1& --286 &68&b \\
520.03& {[}\ion{N}{i}]   &0.0175&49.0& --275 &73&b \\
575.46& {[}\ion{N}{ii}]  &0.0070&19.6& --269 &57&d \\
587.56&  \ion{He}{i}   &0.0069&19.3& --286 &52&d \\
630.03& {[}\ion{O}{i}]   &0.0505&141.5& --298 &85&b \\
636.38& {[}\ion{O}{i}]   &0.0177&49.6& --288 &66&c \\
654.80& {[}\ion{N}{ii}]  &0.1248&349.6& --288 &89&a \\
656.28&  \ion{H}{i}    &0.2778&778.2& --292 &89&a \\
658.34& {[}\ion{N}{ii}]  &0.4120&1154.1& --290 &87&a \\
671.64& {[}\ion{S}{ii}]  &0.1080&302.5& --289 &78&a \\
673.08& {[}\ion{S}{ii}]  &0.0978&273.9& --292 &80&a \\
715.52& {[}\ion{Fe}{ii}] &0.0115&32.2& --295 &66&c \\
729.15& {[}\ion{Ca}{ii}] &0.0442&123.8& --295 &64&c \\
861.70& {[}\ion{Fe}{ii}] &0.0205&57.4& --285 &78&c \\
906.86& {[}\ion{S}{iii}] &0.0064&17.9& --270 &71&d \\
953.06& {[}\ion{S}{iii}] &0.0250&70.0& --266 &97&c \\
982.41& {[}\ion{C}{i}]   &0.0268&75.1& --270 &80&c \\
985.03& {[}\ion{C}{i}]   &0.0952&266.7& --291 &82&b \\
\hline\end{tabular}
\tablefoot{The first two columns identify the lines. The four subsequent columns 
present results of Gaussian fits to  component C. Notation as in Table\,\ref{tab-fluxes-2}.} 
\end{table}

The spectra are heavily reddened. We corrected the spectra for reddening using 
the extinction curve of \citet{draine}.\footnote{available at \url{www.astro.princeton.edu/~draine/dust/dust.html}} To evaluate the amount of extinction, we followed the 
standard procedure often applied to nebular spectra and matched the 
observed Balmer decrements to the theoretical one \citep[cf. Table\,4.4 in][]{Osterbrock}. 
A particular attention was given to reproduce the right ratio of H$\alpha$ 
to H$\beta$ as it is most sensitive to the value of extinction. 
Other Balmer lines are rather faint in the observed spectrum so their measured 
fluxes are subject to larger errors. Table\,\ref{tab-dered} presents the extinction-corrected fluxes  
normalized to H$\beta$ and multiplied by 100. The first row names the spectra 
(e.g. A1 stands for component A in spectrum 1). The second row gives the 
adopted value of the logarithmic extinction, $C$, at H$\beta$ and the 
corresponding value of $E_{B-V}$ in the third row. 

\begin{table*}[]\centering \small
\caption{Intensities of the emission lines corrected for the reddening. The values of the adopted extinction, $C$(H$\beta$), and reddening, $E_{B-V}$, are given for the individual regions in the second and
third row, respectively.}
\label{tab-dered}
\begin{tabular}{lr| rc rc rc rc rc}
\hline\hline
\multicolumn{2}{c}{region:}&\multicolumn{2}{c}{A1}&\multicolumn{2}{c}{A2}&\multicolumn{2}{c}{B1}&\multicolumn{2}{c}{B2}&\multicolumn{2}{c}{C}\\
\multicolumn{2}{c}{$C$(H$\beta$):}&\multicolumn{2}{c}{1.35}&\multicolumn{2}{c}{1.40}&\multicolumn{2}{c}{1.40}&\multicolumn{2}{c}{1.35}&\multicolumn{2}{c}{1.32}\\
\multicolumn{2}{c}{$E_{B-V}$:}&\multicolumn{2}{c}{0.90}&\multicolumn{2}{c}{0.93}&\multicolumn{2}{c}{0.93}&\multicolumn{2}{c}{0.90}&\multicolumn{2}{c}{0.88}\\

\hline 
\multicolumn{1}{c}{Ion}
&\multicolumn{1}{c|}{$\lambda$\,(nm)}&\multicolumn{10}{c}{$I_\lambda/I_{\rm{H}\beta}\cdot 100$ and
its uncertainity flag}\\
\hline
  \ion{H}{i}    & 486.13 &  100.0 &a  &   100.0 &a &    100.0 &a  &   100.0 &a  &   100.0 &a\\
  \ion{H}{i}    & 397.01 &   12.3 &c  &         &  &     13.0 &c  &         &   &         & \\
  \ion{H}{i}    & 410.17 &   27.3 &b  &    30.6 &b &     26.8 &b  &    18.2 &d  &         & \\
  \ion{H}{i}    & 434.05 &   51.3 &b  &    53.8 &b &     49.9 &b  &    49.7 &b  &    56.8 &b\\
  \ion{H}{i}    & 656.28 &  287.0 &a  &   286.4 &a &    281.8 &a  &   287.2 &a  &   282.2 &a\\
  \ion{H}{i}    & 901.49 &    2.0 &c  &     2.0 &c &      1.6 &c  &         &   &         & \\
  \ion{H}{i}    & 922.90 &    2.6 &c  &     2.4 &c &      2.5 &c  &         &   &         & \\
  \ion{H}{i}    & 954.60 &    2.8 &c  &         &  &      2.7 &c  &         &   &         & \\
  \ion{H}{i}    &1004.98 &    6.0 &c  &         &  &      5.7 &c  &         &   &         & \\
  \ion{He}{i}   & 388.86 &   24.0 &c  &    25.0 &b &     25.0 &c  &    53.0 &c  &         & \\
  \ion{He}{i}   & 447.15 &    9.6 &c  &    12.3 &c &     12.3 &c  &         &   &         & \\
  \ion{He}{i}   & 501.57 &    7.6 &c  &         &  &      8.4 &c  &         &   &         & \\
  \ion{He}{i}   & 587.56 &   31.4 &a  &    30.9 &a &     40.8 &a  &    53.2 &b  &     9.9 &d\\
  \ion{He}{i}   & 667.82 &    9.8 &b  &     9.4 &b &      9.7 &b  &    14.4 &c  &         & \\
  \ion{He}{i}   & 706.52 &    5.8 &b  &     5.9 &b &     23.4 &a  &    33.9 &b  &         & \\
  \ion{He}{ii}  & 468.57 &    6.9 &c  &         &  &      7.2 &c  &         &   &         & \\
 {[}\ion{C}{i}]   & 872.71 &        &   &     1.5 &c &          &   &         &   &         & \\
 {[}\ion{C}{i}]   & 982.41 &   12.7 &b  &    19.8 &b &          &   &         &   &    10.7 &c\\
 {[}\ion{C}{i}]   & 985.03 &   32.2 &a  &    56.0 &a &          &   &         &   &    38.0 &b\\
 {[}\ion{N}{i}]   & 519.79 &   54.3 &b  &    71.3 &a &          &   &         &   &    29.7 &b\\
 {[}\ion{N}{i}]   & 520.03 &   65.8 &b  &    77.3 &a &          &   &         &   &    38.2 &b\\
 {[}\ion{N}{ii}]  & 575.46 &   16.6 &b  &    11.5 &c &     17.4 &b  &         &   &    10.7 &d\\
 {[}\ion{N}{ii}]  & 654.80 &  383.2 &a  &   269.0 &a &     91.4 &b  &   153.1 &b  &   127.6 &a\\
 {[}\ion{N}{ii}]  & 658.34 & 1184.9 &a  &   842.4 &a &    273.5 &b  &   318.3 &a  &   414.7 &a\\
 {[}\ion{O}{i}]   & 630.03 &   48.2 &b  &    42.2 &a &          &   &         &   &    58.0 &b\\
 {[}\ion{O}{i}]   & 636.38 &   17.0 &b  &    14.7 &b &          &   &         &   &    19.7 &c\\
 {[}\ion{O}{ii}]  & 372.60 &  145.5 &a  &   112.4 &a &          &   &         &   &   232.5 &b\\
 {[}\ion{O}{ii}]  & 372.88 &  133.5 &a  &   114.3 &a &          &   &         &   &   246.1 &b\\
 {[}\ion{O}{iii}] & 436.32 &        &   &         &  &     17.4 &b  &    18.4 &d  &         & \\
 {[}\ion{O}{iii}] & 495.89 &        &   &     2.5 &d &    311.4 &a  &   533.0 &a  &    21.3 &c\\
 {[}\ion{O}{iii}] & 500.68 &        &   &     7.1 &c &    929.8 &a  &  1583.7 &a  &    60.4 &b\\
 {[}\ion{Ne}{iii}]& 386.88 &        &   &    18.8 &c &    306.0 &a  &   535.1 &a  &         & \\
 {[}\ion{Ne}{iii}]& 396.75 &        &   &     5.7 &d &     69.5 &a  &   141.3 &b  &         & \\
 {[}\ion{S}{ii}]  & 671.64 &  136.4 &a  &   130.0 &a &          &   &         &   &   102.7 &a\\
 {[}\ion{S}{ii}]  & 673.08 &  131.6 &a  &   120.0 &a &          &   &         &   &    92.4 &a\\
 {[}\ion{S}{iii}] & 631.21 &        &   &         &  &      4.8 &b  &         &   &         & \\
 {[}\ion{S}{iii}] & 906.86 &    3.0 &c  &     1.7 &d &     39.4 &a  &    45.2 &a  &     3.0 &d\\
 {[}\ion{S}{iii}] & 953.06 &    7.1 &b  &     5.1 &c &    100.7 &a  &   136.0 &a  &    10.6 &c\\
 {[}\ion{Ar}{iii}]& 713.58 &        &   &         &  &     30.1 &a  &    41.4 &a  &         & \\
 {[}\ion{Fe}{ii}] & 715.52 &   10.1 &b  &     4.9 &b &          &   &         &   &     9.2 &c\\
 {[}\ion{Fe}{ii}] & 717.20 &    1.6 &c  &         &  &          &   &         &   &         & \\
 {[}\ion{Fe}{ii}] & 745.26 &    3.2 &c  &         &  &          &   &         &   &         & \\
 {[}\ion{Fe}{ii}] & 861.70 &   20.4 &b  &    10.2 &b &          &   &         &   &    10.6 &c\\
 {[}\ion{Fe}{ii}] & 889.19 &        &   &     4.1 &b &          &   &         &   &         & \\
 {[}\ion{Fe}{ii}] & 905.20 &    5.5 &b  &         &  &          &   &         &   &         & \\
 {[}\ion{Fe}{ii}] & 922.66 &    4.2 &b  &     1.2 &c &          &   &         &   &         & \\
 {[}\ion{Ni}{ii}] & 737.78 &   13.1 &b  &     5.7 &b &          &   &         &   &         & \\
 {[}\ion{Ca}{ii}] & 729.15 &   27.6 &b  &    18.6 &b &          &   &         &   &    33.6 &c\\
 {[}\ion{Ca}{ii}] & 732.39 &   29.0 &b  &    12.7 &b &          &   &         &   &         & \\
\hline\hline\end{tabular}
\tablefoot{The intensities are expressed relative to H$\beta$ and scaled by 100. They are associated by quality flags a--d which are consistent with the notation in Tables\,\ref{tab-fluxes-1}--\ref{tab-fluxes-3}.} 
\end{table*}

\section{Derivation of $T_{\rm e}$, $N_{\rm e}$, and abundances}\label{sec-analyzis}
Determination of the elemental abundances in CK\,Vul is the primary goal of the present study. 
The theory of line emission processes in nebular conditions is well understood \citep[e.g. Ch.\,4 and 5 in][]{Osterbrock}. There are two kinds of line-excitation mechanisms relevant for ionized nebulae. The lines of \ion{H}{i}, \ion{He}{i}, and \ion{He}{ii} are emitted mainly due to recombination of the H and He ions. The lines of heavier elements, mostly forbidden ones, are produced by collisional excitation of the lowest levels of neutral atoms and ions with free electrons.

We performed the analysis of collisionally excited (forbidden) lines in CK\,Vul by solving the statistical-equilibrium equation for a given electron temperature and density and in a five-level model of a given ion (atom).\footnote{As we argue later on in this paper, the matter in the investigated nebular region of CK\,Vul can be out of the ionization and thermal equilibria. One may think that in that case the assumption of statistical equilibrium is not satisfied as well, affecting the ionic level populations. However,
the processes governing the level populations are much faster than those governing the ionization state. In our case ($N_e \simeq 500$\,cm$^{-3}$ and $T_e \simeq 10^4$\,K), the collisional excitation and de-excitation of low-lying levels of atoms and ions proceeds on a time scale of hours, while the recombination of ionized hydrogen takes hundreds of years. The physical conditions ($N_e$ and $T_e$) in the nebular region of CK\,Vul certainly do not vary on a time scale of hours or days (this region has been observed for decades). Thus, the statistical equilibrium can be safely adopted when calculating the populations of
low-lying levels of atoms and ions.} The necessary radiative probabilities and effective collision strengths
were taken from a compilation\footnote{\url{www.astronomy.ohio-state.edu/~pradhan/atomic.html}} of \citet{pradhan}.

The electron temperature, $T_{\rm e}$, was derived by adjusting the model 
($^1\!S-^1\!D$)/($^1\!D-^3\!P$) line ratio to the observed values of [\ion{C}{i}], [\ion{N}{ii}], [\ion{O}{iii}], and [\ion{S}{iii}]. The same was performed for the ($^2\!D_{3/2}-^4\!S$)/($^2\!D_{5/2}-^4\!S$) line ratio of [\ion{N}{i}], [\ion{O}{ii}], and [\ion{S}{ii}] which yields estimates of the electron density, $N_{\rm e}$. 

The results are given in Table\,\ref{tab-TeNe}. The first part of the table gives our estimates of $T_{\rm e}$. They rely on the quality of the ($^1\!S-^1\!D$) lines, which are often faint in our spectra. Hence the accuracy of the $T_{\rm e}$ determinations, included in Table\,\ref{tab-TeNe}, corresponds to the quality of measurements of this transition. For further analysis, we adopt the most reliable values of $T_{\rm e}$, that is we take a mean value of $T_{\rm e}$([\ion{O}{iii}]) and $T_{\rm e}$([\ion{S}{iii}]) to represent region B1 and $T_{\rm e}$([\ion{N}{ii}]) in  regions A1 and A2. In region B2 we have no reliable estimate of $T_{\rm e}$. Since the spectrum of this region shows a similar degree of excitation as that of region B1, we adopted for region B2 the same $T_{\rm e}$ as for region B1, i.e. $T_{\rm e}$=15000\,K. Region C, where we cannot reliably  estimate $T_{\rm e}$, exhibits similar excitation as regions A1 and A2. Therefore, we adopted $T_{\rm e}$=10000\,K for region C.

\begin{table*}[]\centering \small
\caption{Physical conditions, $T_{\rm e}$ and $N_{\rm e}$, and abundances derived for the individual regions.}
\label{tab-TeNe}
\begin{tabular}{lcccccccccc}
\hline\hline
\multicolumn{1}{c}{region}&\multicolumn{2}{c}{A1}&\multicolumn{2}{c}{A2}&\multicolumn{2}{c}{B1}&\multicolumn{2}{c}{B2}&\multicolumn{2}{c}{C}\\ 
\hline
&\multicolumn{10}{l}{$T_{\rm e}$ (K)}\\
\cline{2-11} 
 {[}\ion{O}{iii}] &        &   &        &   & 14800 &b & 12200 &d &         &  \\
 {[}\ion{S}{iii}] &        &   &        &   & 15000 &b &         &  &         &  \\ 
 {[}\ion{N}{ii}]  & 10100  & b & 10100& c& 26500 &b &         &  & 13700 &d \\
 {[}\ion{C}{i}]   &        &   &  6310& c&       &  &         &  &         &  \\  
{\bf adopted} & {\bf 10100}&   & {\bf 10100}&   & {\bf 14900} &  & {\bf 15000} &  & {\bf 10000} &  \\
\hline 
&\multicolumn{10}{l}{$N_{\rm e}$ (cm$^{-3}$)}\\
\cline{2-11} 
 {[}\ion{O}{ii}]   &    585 & a &    417 & a  &        &&            &&       357 & b \\
 {[}\ion{S}{ii}]   &    503 & a &    414 & a  &        &&            &&       364 & a \\
 {[}\ion{N}{i}]    &    188 & b &    323 & a  &        &&            &&       129 & b \\
{\bf adopted} & {\bf 544} &   &{\bf 415}&    &{\bf 600}  &&{\bf 200}&& {\bf 360} &   \\
\hline 
&\multicolumn{10}{l}{ionic abundances}\\
\cline{2-11} 
 N${^+}$/H${^+}$   & 2.17E--4 &a  &1.57E--4 &a  &2.12E--5 &b  &2.69E--5 &a  &7.79E--5 &a\\
 O${^+}$/H${^+}$   & 1.08E--4 &a  &8.83E--5 &a  &         &   &         &   &1.90E--4 &b\\
O$^{++}$/H${^+}$   &          &   &2.39E--6 &c  &9.89E--5 &a  &1.66E--4 &a  &2.10E--5 &b\\
Ne$^{++}$/H${^+}$  &          &   &1.49E--5 &d  &5.94E--5 &a  &1.05E--4 &b  &         & \\
 S${^+}$/H${^+}$   & 5.87E--6 &a  &5.42E--6 &a  &         &   &         &   &4.26E--6 &a\\
S$^{++}$/H${^+}$   & 3.53E--7 &b  &2.44E--7 &c  &2.55E--6 &a  &3.27E--6 &a  &4.90E--7 &c\\
Ar$^{++}$/H${^+}$  &          &   &         &   &1.21E--6 &a  &1.64E--6 &a  &         & \\
\hline  
&\multicolumn{10}{l}{elemental abundances}\\
\cline{2-11} 
 He/H   &  0.229     &&  0.228      && 0.259      && 0.391     &&           \\             
 N/H    &  2.17E--4  &&  1.61E--4   &&            &&           &&  8.65E--5 \\             
 O/H    &  1.08E--4  &&  9.07E--5   && 9.89E--5   && 1.66E--4  &&  2.11E--4 \\             
 Ne/H   &            &&             && 5.94E--5   && 1.05E--4  &&           \\             
 S/H    &  6.22E--6  &&  5.67E--6   && 2.55E--6   && 3.27E--6  &&  4.75E--6 \\             
 Ar/H   &            &&             && 1.21E--6   && 1.64E--6  &&           \\             
\hline\end{tabular}
\tablefoot{The letters next to the physical quantities indicate the quality of the original flux measurements of the relevant
lines: a, flux accurate to within 10\%; b, error between 10 and 30\%; c, error greater than 30\%; d, line marginally detected.} 
\end{table*}

The second part of Table\,\ref{tab-TeNe} presents our estimates of $N_{\rm e}$. 
For further analysis of regions A1, A2, and C, we adopted mean values of 
$N_{\rm e}$ derived from the line ratios of [\ion{O}{ii}] and [\ion{S}{ii}]. 
In the case of regions B1 and B2, no measurements of line ratios sensitive to 
$N_{\rm e}$ are available. Instead, we estimated $N_{\rm e}$ from comparing 
the surface brightness of H$\beta$ in regions A and B. Note that the 
emissivity in H$\beta$ is proportional to $N_{\rm e}^2$. As can be seen in 
Fig.\,\ref{fig-prof1}, H$\beta$ is slightly stronger in region B1 than in 
region A1. Therefore, we adopted $N_{\rm e}$=600\,cm$^{-3}$ for region B1. 
Similarly, comparing the H$\beta$ flux in region B2 to that in A2 
(see Fig.\,\ref{fig-prof2}), we adopted $N_{\rm e}$=200\,cm$^{-3}$ for B2. 
Note that the ion abundances are rather insensitive to the actual value 
of $N_{\rm e}$ in the range relevant for our spectra, i.e., 
for $N_{\rm e} \lesssim 10^4$\,cm$^{-3}$ (see also below).

As can be seen in Table\,\ref{tab-TeNe}, the [\ion{N}{ii}] ratio in region B1 yields $T_{\rm e}$=26500\,K. This is much higher than the temperature derived from the line ratio of [\ion{O}{iii}] or [\ion{S}{iii}]. The discrepancy may suggest that the density in region B1 is actually higher than the adopted value of 600\,cm$^{-3}$. Indeed, $T_{\rm e}$ derived from the [\ion{N}{ii}] ratio is more sensitive to $N_{\rm e}$ than in cases of [\ion{O}{iii}] and [\ion{S}{iii}], especially at higher $N_{\rm e}$. Our calculations show that at $N_{\rm e}$=$4\times 10^4$\,cm$^{-3}$ the three line ratios would indicate similar temperatures of about 13800\,K. At this high density, however, the emissivity of H$\beta$ in region B1 would be almost $10^4$ times higher than that in region A1, whereas the measured fluxes of H$\beta$ in regions A1 and B1 are comparable (Table\,\ref{tab-fluxes-1}). This implies that the volume occupied by nebular matter in region B1 is $\sim\!10^4$ times smaller than in region A1. Yet, the extent along the slit and in the range of radial velocities of component B are similar to that of component A (Fig.\,\ref{fig-maps}). The only way to reconcile these contradictions is to propose a two-phase structure of region B1. Namely that the bulk of the emission comes from a thin ($N_e\approx 600$\,cm$^3$) medium of high excitation (doubly ionized O, Ne, S, and Ar), in which there are small but dense inclusions of low ionization that give rise to the [\ion{N}{ii}] emission.

Having determined $T_{\rm e}$ and $N_{\rm e}$, we calculated the emissivity of 
H$\beta$ and of the forbidden lines, see e.g. Ch.\,5.11 in \citet{Osterbrock}. Comparing 
the emissivity in a given line normalized to H$\beta$ with the corresponding 
measured intensities, we obtained the ion abundances relative to ionized 
hydrogen. The results are given in Table\,\ref{tab-TeNe}.

We next discuss uncertainties in the derived
abundances. The accuracy of ionic abundances depend mainly on the
precision of measurements of the line intensities. When calculating the ionic abundances, we used
the strongest lines of a given ion, i.e. the lines whose fluxes
were most accurately measured. For example, the abundance of 
O$^{++}$ was obtained from the sum of the flux of [\ion{O}{III}] 495.9 and
500.7. Thus, at this stage, the derived values are accurate to a few percent.
However, the line emissivities depend not only on the ion abundances but
also on the electron density and temperature, $N_e$ and $T_e$. The
derivation of these parameters were often based on weaker lines,
and this can be a source of larger uncertainties in the final
abundances.

$N_e$ was mainly estimated from the line ratios of [\ion{O}{II}] and
[\ion{S}{II}]. These
intensities were usually quite precisely measured (quality flag a in Tables\,\ref{tab-fluxes-1}--\ref{tab-fluxes-3}). Our calculations show that a 10\%\ uncertainty
in these line ratios results in a 30--40\%\ uncertainty in $N_e$. Although, the error is
quite large, it has a negligible effect on the final
ionic abundances, as the resulting uncertainty in the abundances 
never exceeds 4\%. 

$T_e$ was calculated from the line
ratios of [\ion{N}{II}], [\ion{O}{III}], and [\ion{S}{III}]. 
These ratios involve the emission from
the ($^1\!S - ^1\!D$) transition, which is usually quite faint (quality flag b, at
best). Calculations show that a 20\%\ uncertainty in the involved line
ratios results in a 7 -- 10\%\ uncertainty in $T_e$. However, this small
error in $T_e$ causes much larger errors in the abundances since the line
emissivities depend exponentially on $T_e$. Because of this exponential 
dependence the error of ionic abundances also depends on the excitation
energy of the transition, in other words, on the wavelength of the
line used to derive the abundance. The shorter the wavelength, the
larger the error. In our case, the final
uncertainties in the ionic abundances range from 16\%\ for S$^{++}$, 
$\sim$20\%\ for S$^+$, N$^+$ and Ar$^{++}$, 28\%\ for O$^{++}$, up to 
$\sim$35\%\ for Ne$^{++}$ and O$^+$.

Not every ion of a given element is observable in the wavelength range covered 
by our observations. This has to be taken into account when deriving 
the abundances of the elements. The standard way is to apply the
so-called ionization correction factors, usually calculated from
steady-state photoionization models \citep[e.g.][]{stas}. However, it would
be somewhat risky in our case, as
it is not clear what the primary ionization mechanism is operating in the nebular regions 
in CK\,Vul.
In Sect.\,\ref{sec-shocks}, we argue that these are shocks rather than
photoionization by the central star. In the case of shocks the optical
emission lines are mainly produced in ambient matter photoionized by
radiation emitted in the hot post-shock region and/or in matter recombining
from the hot, highly ionized post-shock conditions. Finally, at the observed
densities ($N_e \approx$ 500\,cm$^{-3}$) the recombination time is long
($\sim$250 years for H$^+$). Therefore we can also deal with
matter that was ionized in the past and is now recombining.
The state of the latter can be far from ionization equilibrium conditions.
Therefore, we follow a more general reasoning that allows us to get reliable 
estimates of the total elemental abundances, independent of the actual ionization 
mechanism and history.

The low flux of the \ion{He}{ii} 468.6 line assures that the abundance of 
He$^{++}$ is negligible in all the investigated regions. Indeed, 
our measurements of this line (see Table\,\ref{tab-dered}) yield a
ratio He$^{++}$/H$^+ \lesssim$ 0.005. We thus neglect 
the \ion{He}{ii} 468.6 line while determining the He abundance.
Consequently, the abundance of helium was derived from the intensities of 
the \ion{He}{i} lines only. We used \ion{He}{i} emissivities of \cite{Benjamin}. 
Their formulas, besides the recombination cascades, take into account 
collisional effects. The lines \ion{He}{i} 388.9 and 706.5 can be subject to 
radiative-transfer effects \citep[see e.g. Ch.\,4.6 in][]{Osterbrock}, so we excluded
them from our abundance calculations. The final He abundance was obtained 
as a weighted mean of values derived from each individual line. 
The weights were assigned according to the precision of the original flux 
measurement. 
The results are given in Table\,\ref{tab-TeNe}. Uncertainties
of the listed He abundances do not exceed 15\%.

The negligible abundance of He$^{++}$ allows us to assume that all ions with 
an ionization potential $\geq$54.4\,eV are practically absent. 
This concerns the following (or higher) ions of interest: N$^{4+}$, 
O$^{3+}$, Ne$^{3+}$, S$^{5+}$, and Ar$^{4+}$. All these ions have
recombination coefficients much larger than that of He$^{++}$ and very fast 
rates of charge transfer with H$^0$ \citep{kf96}. Therefore if the
ionization processes cannot maintain a substantial abundance of He$^{++}$,
they have even more difficulties with making the above ions abundant.

Thus, in the case of oxygen, all the presumably abundant ions, i.e. O$^+$ and
O$^{++}$, are observed and we assume 
$$       \rm{O/H} = (\rm{O}^+/{\rm H}^+) + ({\rm O}^{++}/H^+).$$

In the case of nitrogen, only N$^+$ is observed, while N$^{++}$ and N$^{3+}$
can be potentially abundant. In component B, the latter ions are probably
dominant. Therefore we cannot reliably estimate the N abundance in region B.
In component A (and C), the [\ion{N}{II}] lines are very strong. This suggests
that N$^+$ is the dominant ion of N. This conviction is strengthened by
the weakness of [\ion{O}{III}] lines and consequently the very low abundance
of O$^{++}$ in component A. We adopted the following formula
$$       \rm{N/H} = (\rm{N}^+/\rm{H}^+)(\rm{O}^+/\rm{H}^+)/(\rm{O/H})$$
when determining the N abundance. This is a standard ionization correction
formula used in the case of photoionized nebulae. In our case of component A, 
it practically assumes that the N abundance is equal to the observed abundance 
of N$^+$. If N$^{++}$ is substantially abundant, then we underestimate the
N abundance.

Sulfur is the second element, after oxygen, whose two ionization states,
i.e. S$^+$ and S$^{++}$, are observed in CK\,Vul. However, contrary to O, 
we cannot simply assume that these are the only ions of S that are abundant,
as the presence of S$^{3+}$, or even S$^{4+}$, cannot be excluded. 
This is particularly relevant for component B where the
[\ion{S}{III}] lines are strong and those of [\ion{S}{II}] are not observed.
Nevertheless, we adopt
$$       \rm{S/H} = (\rm{S}^+/\rm{H}^+) + (\rm{S}^{++}/\rm{H}^+)$$
as nothing better can be done at this time. We expect this is a good approximation in component A (and C), as S$^{++}$ is there at least an order of magnitude less abundant
than S$^+$ and therefore the abundances of S$^{3+}$ and S$^{4+}$ should be negligible. 
The latter is not expected in component B and thus our procedure yields an underestimated abundance of S in region
B. This systematic error has a confirmation in our final results for all the regions in Table\,\ref{tab-TeNe} where S/H in component B is systematically lower than that in component A and C. Such discrepancy is not expected if the relative elemental abundances are constant throughout the small part of the nebula probed with the X-shooter slit.

Neon and argon are observed only in component B and only in the form of doubly
ionized species. In the case of neon, Ne$^{++}$ is most probably the dominant ion
in component B. Ne$^+$ should be negligible, because O$^+$ is not observed. 
Admittedly, Ne$^+$ has a higher ionization potential than O$^+$ but 
a very low rate of charge transfer between Ne$^{++}$ and H$^0$,
compared to that between O$^{++}$ and H$^0$ \citep{kf96}, should 
compensate for the effect caused by the difference in ionization potentials. 
Therefore, we simply assume
$$       \rm{Ne/H} = \rm{Ne}^{++}/\rm{H}^+.$$

In the case of argon, Ar$^{3+}$ cannot be excluded as an important ion in component B. 
Therefore, a determination
of Ar abundance solely from the measured abundance of Ar$^{++}$, i.e.
$$       \rm{Ar/H} = \rm{Ar}^{++}/\rm{H}^+,$$
leads to an underestimated abundance of this element.

With the elemental abundances obtained for individual regions in 
Table\,\ref{tab-TeNe}, we derived the mean abundances representing all the 
regions. When averaging, the results from regions B2 and C were weighted 
twice lower than those from the other regions. The spectra from these two 
regions are significantly fainter and, consequently, the measurements of line 
intensities were less precise than those in the other regions. As a result, 
we were unable to derive reliable estimates of the electron
temperature, which introduced an additional uncertainty in the abundances 
for these two regions. When calculating the mean abundance of
sulfur, we omitted the results obtained for component B, because they probably
underestimate the S/H value (see above).
The upper part of Table\,\ref{tab-final} lists the mean number abundances
and their estimated errors.
The bottom part of Table\,\ref{tab-final} presents the same abundances but 
in mass fractions. 
For comparison, we also listed the abundances of the Sun from \cite{Asplund}.

\begin{table}[]\centering \small
\caption{Mean elemental abundances of CK\,Vul compared to solar ones.}
\label{tab-final}
\begin{tabular}{cccc}
\hline\hline
\multicolumn{3}{c}{relative abundances by number:}\\
\hline
        &   CK\,Vul  & error &    Sun       \\ \cline{2-3}
 He/H   &   0.260    & $\pm$21\% &  0.085       \\
 N/H    &   1.69E--4 & $\pm$29\% &  6.76E--5    \\
 O/H    &   1.22E--4 & $\pm$34\% &  4.90E--4    \\
 Ne/H   &   7.45E--5 & $\pm$29\% &  8.50E--5    \\
 S/H    &   5.70E--6 & $\pm$~~9\% &  1.32E--5    \\
 Ar/H   &   1.35E--6 & $\pm$15\% &  2.50E--6    \\
\hline
\multicolumn{3}{c}{abundances by mass:}\\
\hline
     &   CK\,Vul   & error &    Sun      \\ \cline{2-3}
 H   &   0.488     & $\pm$18\% &  0.740      \\
 He  &   0.509     & $\pm$28\% &  0.252      \\
 N   &   1.15E--3  & $\pm$35\% &  7.01E--4   \\
 O   &   9.50E--4  & $\pm$40\% &  5.80E--3   \\
 Ne  &   7.27E--4  & $\pm$35\% &  1.26E--3   \\
 S   &   8.91E--5  & $\pm$16\% &  3.13E--4   \\
 Ar  &   2.64E--5  & $\pm$21\% &  7.40E--5   \\
\hline\hline
\end{tabular}
\end{table}

\section{Discussion}\label{sec-discuss}

\subsection{The excitation mechanism of the nebula}\label{sec-shocks}
There are two possible mechanisms producing nebular line emission: direct ionization by stellar radiation and excitation by shock fronts. It is not straightforward to distinguish between these two mechanisms from observations of line intensities alone. This is because in the case of shocks, the nebular lines, especially in the optical, are excited mostly through ionization by radiation generated in the shock. Both mechanisms are essentially incarnations of photoionization. In the case of CK\,Vul, there are however several arguments favoring shocks instead of photoionization by a hot central star. 

First, we do observe gas moving at high velocities. As can be seen in Fig.\,\ref{fig-maps}, the observed radial velocities range from $\sim$--300 to $\sim$+350\,\kms. The de-projected gas velocities are probably much higher. Indeed, the large-scale hourglass optical nebula studied by \citet{hajduk2013} appears to have a large inclination of about 65\degr\ at which the gas moves mainly perpendicular to the line of sight and the tips of that nebula are estimated to move as fast as 900\,\kms. Dissipation in shocks with velocities of a few hundreds of \kms\ can  result in gas temperatures above $10^6$\,K and produce intense ultraviolet and X-ray radiation capable of photoionizing large volumes of the surrounding medium \citep{dopita96}. 

Second, the emission lines of neutral CNO elements (that is, [\ion{C}{i}], 
[\ion{N}{i}], and [\ion{O}{i}]) and singly ionized metals ([\ion{Fe}{ii}], 
[\ion{Ni}{ii}], and [\ion{Ca}{ii}]) are particularly strong in the spectrum 
of CK\,Vul. This is generally considered as a signature of shock-excited emission regions \citep[e.g.][]{dopita95}. Shocks, contrary to photoionizantion by a star, produce extended recombination regions of a low ionization degree, where hydrogen and elements having similar ionization potential as hydrogen -- such as N
and O -- remain largely neutral, while elements of low-ionization potentials -- including, Fe, Ni, and Ca -- are predominantly singly ionized \citep[e.g.][]{dopita96}.

Third, we do not observe any significant variation of the ionization
and excitation conditions with the radial distance from the position of the
presumable central object. Regions A2 and B2 are significantly farther away 
from the location of the central star than regions A1 and B1, 
respectively. Yet, comparing regions A1 with A2 and B1 with B2, 
their spectra (Table\,\ref{tab-dered}) as well as the ionization degrees and 
temperatures (Table\,\ref{tab-TeNe}) are very similar. We do, however, observe 
that the regions at the same (projected) distance from the central star
but at different radial velocities, i.e. A1 and B1 or A2 and B2, 
have drastically different spectra and ionization 
states. In the case of photoionization by a star, 
this would imply that the star emits drastically different spectra in 
different directions, which is unlikely. The differences between components A 
and B can, however, be easily understood, if they are powered by shocks.
The lower ionization and excitation degree of component A may imply that 
the matter at negative radial velocities was shock-excited some time before
the matter of component B (at positive radial velocities). Component A 
had thus more time to recombine and cool off than component B. We note that 
at the derived electron densities ($\sim$500\,cm$^{-3}$) the recombination 
time of ionized  hydrogen is $\sim$250 years. Ions, like O$^{++}$ and 
Ne$^{++}$, recombine much faster than H$^+$.

Another way of explaining the differences between component A and B in terms
of shocks is to assume that component B was excited by a stronger shock than
component A. In the case of low-velocity shocks ($\varv_{\rm s} \lesssim$
150\,km\,s$^{-1}$), the Balmer decrement is steeper than that of the recombination
case B \citep{dopita17}. This is because collisional excitation 
of H$\alpha$ is important in an extended region of partial ionization
generated by slow shocks.
This seems to be observed in CK\,Vul, particularly in components A and C.
When constraining the excitation conditions in Sect.\,\ref{sec-intens}, we paid attention 
to reproduce the case B ratio of H$\alpha$ to H$\beta$ as this ratio is best determined observationally and is most sensitive to extinction. The results in Table\,\ref{tab-dered} were
obtained under this assumption. However, for all components in this table, the H$\gamma$ relative line intensity is greater than the value of $\sim$47 corresponding to pure recombination.
The excess ranges from $\sim$6\%\ in component B to $\sim$20\%\ in component C.
Formally, its magnitude is within the measurement errors but it may be significant because it appears in all the regions and is correlated with the excitation and ionization state of the components.

We therefore made an exercise and derived the extinction value from
H$\gamma$/H$\beta$. This ratio is supposed to be much less affected by
collisional excitation than H$\alpha$/H$\beta$. We obtained
C(H$\beta$) equal to 0.75 for component C, 1.00--1.08 for component A, and
1.22--1.25 for component B. These values can be compared to those in
Table\,\ref{tab-dered}.
After correcting the observations using these new values of extinction, 
we obtained the relative H$\alpha$ intensity ranging from $\sim$317 for component B, 
353--389 for component A, and 437 for component C.
According to \citet{dopita17}, these values would
correspond to shock velocities of $\sim$35\,km\,s$^{-1}$ for 
component C, $\sim$60\,km\,s$^{-1}$ for component A, and 
$\sim$100\,km\,s$^{-1}$ for component B.
A detailed comparison of the
models of \citet{dopita17} with our spectra is beyond the scope of the
present paper. In fact, it would not be straightforward because
of significant differences in the elemental abundances. 
Moreover, our measurements probe only a narrow cut through the nebular
region of CK\,Vul and probably encompass only a part of a shock-excited region.
We are unable to state which particular parts of the
shock-excited region were observed (e.g. a precursor or postshock flow as in Sutherland \& Dopita) whereas such a distinction would be crucial for comparing
our line intensities with the models.
Nevertheless, the above estimates on $\varv_{\rm s}$ increase with the increasing 
excitation of the observed spectra, as expected from the models. For
instance, the
models predict significant emission in the [\ion{O}{III}] lines only for 
$\varv_{\rm s} \gtrsim$ 80\,km\,s$^{-1}$; that is consistent with the observation of [\ion{O}{III}] in component B only.

It should, however, be emphasized that the above arguments in favor of
shock excitation are not fully conclusive. 
Any information on the effective temperature or 
spectral type of the central object would be more
decisive in this matter. Unfortunately nothing of this kind can be derived
from the observational data currently available for CK\,Vul.

\subsection{The abundances and nature of CK\,Vul}
The elemental abundances in the nebular
region of CK\,Vul presented in Table\,\ref{tab-final} are the primary result of this study. 
They are based on the spectra corrected for the extinction derived in a standard way, i.e. 
assuming that the true Balmer decrement is that of the recombination case B.
In Sect.\,\ref{sec-shocks}, we discussed the possibility of excitation by slow shocks,
which produce the H$\alpha$/H$\beta$ ratio well above the classical 
recombination value. Shock excitation results in extinction values that are significantly
lower than those derived assuming the pure recombination decrement. In order to derive abundances in the framework of slow shocks, the line intensities should be recalculated with the relevant extinction.

We therefore repeated the abundance determination using the extinction values derived
from the observed ratio of H$\gamma$/H$\beta$ (see Sect.\,\ref{sec-shocks}).
For component B, the resultant abundances are close to those in
Table\,\ref{tab-TeNe}. Larger differences are noticeable for components A and
C. The new abundances are more
consistent with each other than those in Table\,\ref{tab-TeNe}. As a result, the uncertainties of the
final averaged abundances, presented in Table\,\ref{tab-final2}, 
are smaller than those in Table\,\ref{tab-final}.

\begin{table}[]\centering \small
\caption{Mean elemental abundances of CK\,Vul derived from the spectra
corrected for extinction using the observed H$\gamma$/H$\beta$ (see Sect.\,\ref{sec-shocks}).}
\label{tab-final2}
\begin{tabular}{ccc}
\hline
\multicolumn{3}{c}{abundances by mass:}\\
\hline
     &   CK\,Vul   & error       \\ \cline{2-3}
 H   &   0.458     & $\pm$17\%       \\
 He  &   0.538     & $\pm$24\%      \\
 N   &   1.59E--3  & $\pm$28\%   \\
 O   &   8.30E--4  & $\pm$30\% \\
 Ne  &   6.70E--4  & $\pm$28\% \\
 S   &   1.26E--4  & $\pm$10\% \\
 Ar  &   2.98E--5  & $\pm$15\% \\
\hline\hline
\end{tabular}
\end{table}

\begin{figure}
\centering
\includegraphics[width=0.5\textwidth]{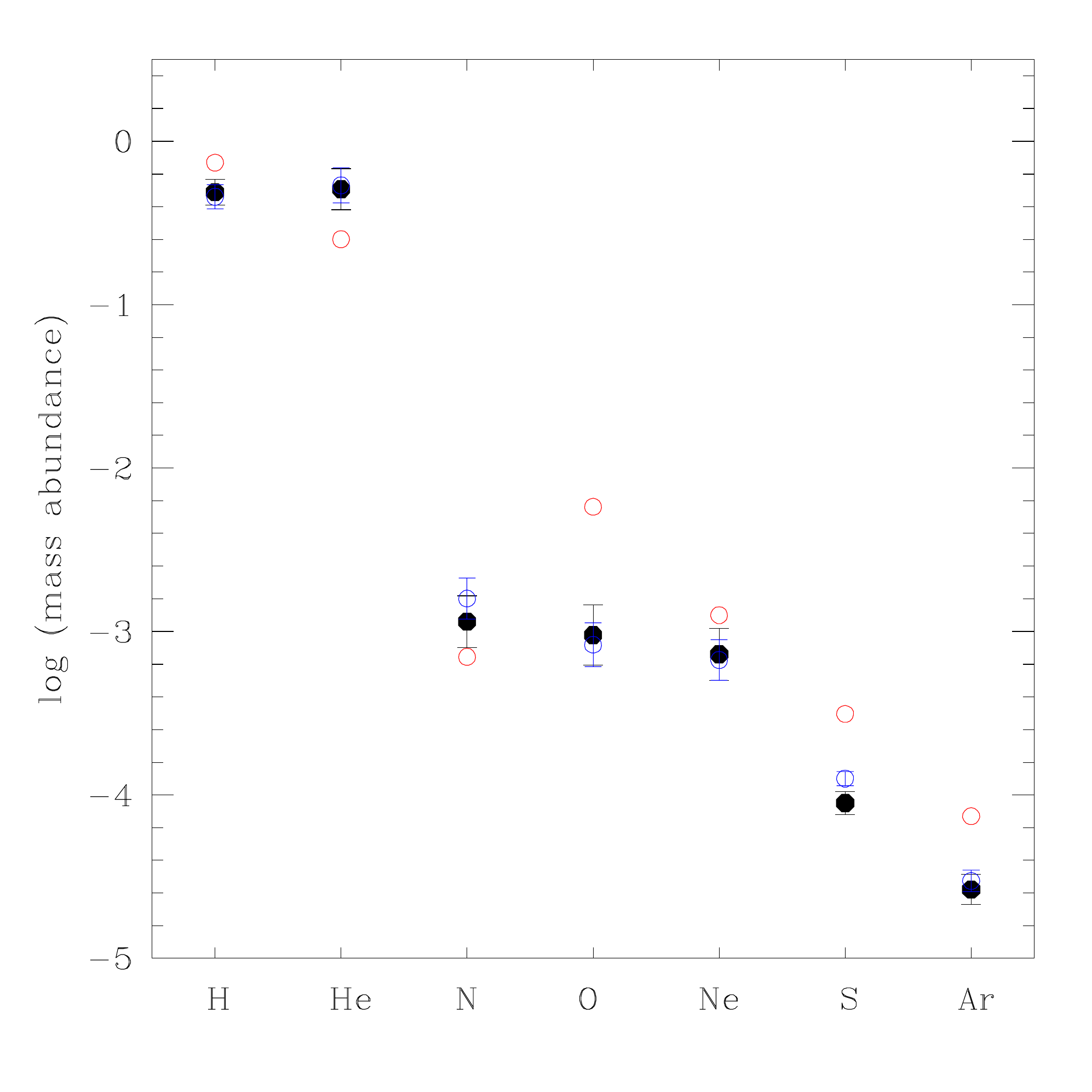}
\caption{Elemental mass abundances in CK\,Vul. Black dots: abundances from
Table\,\ref{tab-final}. Blue circles: abundances 
obtained from the spectra corrected for extinction using the observed
H$\gamma$/H$\beta$ ratio, as explained in Sect.\,\ref{sec-shocks}. Red
circles: abundances in the sun.}
\label{fig-abund}
\end{figure}

In Fig.\,\ref{fig-abund}, we compare the elemental abundances derived for classical case B (from Table\,\ref{tab-final}) and in the presence of shocks (from Table\,\ref{tab-final2}). The
differences are insignificant. Note, however, that the N/O
ratio increases from $\sim$1.2 in Table\,\ref{tab-final} (extinction derived from H$\alpha$/H$\beta$) to $\sim$1.9 in
Table\,\ref{tab-final2} (extinction from H$\gamma$/H$\beta$).

In any case, the abundances we obtained for CK\,Vul are significantly different from 
the solar values. Helium is overabundant by a factor of two, which implies
that $\sim$25\% of hydrogen in the observed matter has been converted to helium. 
As a result, 
the mass abundance of helium is comparable to that of hydrogen. Unfortunately, 
we were not able to derive the carbon abundance, but the obtained abundances 
of nitrogen and oxygen show that a significant part of the CNO elements were 
converted to nitrogen. This is apparent when comparing the N/O ratio, which 
is $\sim$1.2--1.9 in CK\,Vul, with the much lower solar value of 0.12. 
The latter result confirms the suggestion of \citet{kamiNat,kamiIRAM}, based 
on submm spectral analysis, that the remnant of CK\,Vul is rich in nitrogen. 
Overall, the current study strongly reinforces the conclusion drawn from 
the isotopic study of molecular lines that the remnant's material has been 
processed in H-burning in CNO cycles. Possible scenarios for this processing 
were discussed in \citet{kamiIRAM} \citep[see also][]{kami26Al}. 

Table\,\ref{tab-final} also shows that neon, sulfur, 
and argon are underabundant in CK\,Vul compared to the Sun. The abundances 
of these elements are not affected in the CNO cycles. As can also be seen 
in Table\,\ref{tab-final}, the summary abundance of oxygen and nitrogen in CK\,Vul is three times lower than in the Sun. The CNO cycles do not alter the summary abundance of the CNO elements but mainly convert carbon and oxygen to nitrogen. Therefore, we may conclude that elements heavier than helium were less abundant in the matter from which the progenitor of CK\,Vul  was formed than in the Sun. Using the $\zeta$-scaling of elemental abundances with metallicity of \citet{metal}\footnote{see \url{https://miocene.anu.edu.au/mappings/abund/}}, for Ne, S and Ar we find [Fe/H]=--0.53$\pm$0.07. This low-metallicity chemical composition strongly suggests that the progenitor of CK\,Vul was older than the Sun.

\subsection{Extinction}\label{sec-extin}
The observed spectrum of CK\,Vul is significantly reddened. This is mainly
due to interstellar extinction but a contribution from internal dust in
CK\,Vul is not excluded. The observed nebular region ("jet") is situated inside the
northern lobe seen in the submillimeter dust-emission
\citep{kami26Al,evans2018}. Therefore it is likely that part
of the observed reddening is due to local dust situated \emph{outside} the 
observed nebular region but crossed by the line of sight. 
High interstellar extinction towards CK\,Vul is expected based 
on its low Galactic latitude of 1\degr\ and on observations of interstellar 
CO lines \citep{kamiIRAM}. 

The values of extinction listed in Table\,\ref{tab-dered} are practically
the same for all the investigated regions, i.e. $C$(H$\beta$) $\simeq$ 1.36 
($E_{B-V} \simeq$ 0.91). This could suggest that there is no significant
reddening inside the nebular region itself. However, adopting the hypothesis
of slow shocks and determining
the extinction from the observed H$\gamma$/H$\beta$ ratio instead of H$\alpha$/H$\beta$ (see
Sect.\,\ref{sec-shocks}), we derived 
extinction values which are lower and significantly different for 
different regions: that is, 
$C$(H$\beta$) $\simeq$ 0.75 for component C, $\sim$1.04 for component A, and
$\sim$1.24 for component B. We note that the derived $C$
values correlate with the mean radial velocities of the regions (see
Sect.\,\ref{sec-intens}), i.e. --285\,km\,s$^{-1}$ for component C,
$\sim$--130\,km\,s$^{-1}$ for component A, and $\sim$+75\,km\,s$^{-1}$ for
component B. The regions of negative radial
velocities are expected to be closer to us than the one moving with the
positive velocity. Therefore, in this case, we would have to conclude that
the extinction within the nebular region is significant and that the
extinction obtained for component C, $C$(H$\beta$) $\simeq$ 0.75, is an
upper limit to the interstellar extinction.

\section{Summary}\label{sec-summ}

We obtained an optical spectrum of the brightest nebular region of CK\,Vul, 
a remnant of an ancient stellar explosions observed in 1670--72. The spectrum 
was recorded with the X-shooter spectrograph at the VLT and reveals a plethora 
of ionic emission lines typical for spectra of planetary nebulae. An analysis 
of the spectrum indicates the presence of several spatio-kinematical regions 
located $\sim$1\farcs3--8\farcs5 from the presumed position of the star. 
From the measured line intensities, we estimated the electron temperatures, 
typically of 10--15\,kK, and densities of 200--600\,cm$^{-3}$ prevailing in 
the individual regions. This allows us to derive abundances of the ions 
emitting the observed lines and also the abundances of several key elements, 
including H, He, O, and N. Despite the different excitation conditions within 
the nebula, the elemental abundances derived for individual regions are 
very similar. We found that helium is overabundant by a factor of two 
compared to the Sun. Also, nitrogen is overabundant. This is particularly 
well seen in the N/O ratio which is ten times higher than the solar one. 
These findings clearly show that the matter in the remnant of CK\,Vul has been 
processed in the CNO cycles of H-burning. Other elements, such as neon, sulfur, 
and argon, are underabundant compared to the Sun. We have found that the 
overall abundance of the CNO elements in CK\,Vul is also lower than in matter 
of solar (cosmic) composition. Clearly, the progenitor of CK\,Vul was 
an old system, much older than the Sun. The derived abundances together 
with other recent observational results will be discussed in a separate paper 
devoted entirely to revealing the nature of CK\,Vul and its progenitor. 
Finally, we argue that the nebular regions in CK\,Vul are excited by shocks 
rather than being directly photoionized by a central star.

\begin{acknowledgements}
We thank our referee and editor, S. Shore, for comments on the manuscript. 
We are grateful to M. Hajduk for making available to us his processed Gemini image of CK\,Vul.
Based on observations collected at the European Organisation for 
Astronomical Research in the Southern Hemisphere under ESO programme 
099.D-0010(A). R.T. acknowledges a support from grant 
2017/27/B/ST9/01128 financed by the Polish National Science Centre. 
\end{acknowledgements}

\end{document}